\documentclass[useAMS,usenatbib]{mn2e}
\usepackage{graphicx}
\usepackage{epsfig} 
\usepackage{multirow} 
\usepackage{aas_macros}

\def\gsim{\;\rlap{\lower 2.5pt \hbox{$\sim$}}\raise 1.5pt\hbox{$>$}\;} 
\def\lsim{\;\rlap{\lower 2.5pt \hbox{$\sim$}}\raise 1.5pt\hbox{$<$}\;}

\newcommand{\kms}{{\rm \, km \, s^{-1}}}
\newcommand{\msun}{\rm{M}_{\odot}}

\include{mydefs}
\title[Stellar Disks in Aquarius Dark Matter Haloes]
{Stellar Disks in Aquarius Dark Matter Haloes}

\author[J. DeBuhr, C.-P. Ma and S. White]
{Jackson DeBuhr$^{1}$,Chung-Pei Ma$^{1}$, Simon D. M. White$^{2}$ \\
$^{1}$Department of Astronomy, University of California, Berkeley, CA 94720, USA \\
$^{2}$Max-Planck-Institut f\"{u}r Astrophysik, Karl-Schwarzschild-Str. 1, 85748 Garching, Germany}

\begin{document}

\pagerange{ 
\pageref{firstpage}-- 
\pageref{lastpage}} 

\maketitle

\label{firstpage} 
\begin{abstract}
  We investigate the gravitational interactions between live stellar 
  disks and their dark matter halos, using $\Lambda$CDM
  haloes similar in mass to that of the Milky Way taken from the
  Aquarius Project. We introduce the stellar disks by first
  allowing the haloes to respond to the influence
  of a growing rigid disk potential from $z = 1.3$ to $z = 1.0$.  The rigid
  potential is then replaced with star particles which evolve
  self-consistently with the dark matter particles until $z = 0.0$.  Regardless of
  the initial orientation of the disk, the inner parts of the haloes
  contract and change from prolate to oblate as the
  disk grows to its full size.  When the disk normal is initially
  aligned with the major axis of the halo at $z=1.3$, the length of the
  major axis contracts and becomes the minor axis by $z=1.0$.  Six out of
  the eight disks in our main set of simulations form bars,
  and five of the six bars experience a buckling instability that results in
  a sudden jump in the vertical stellar velocity dispersion and an
  accompanying drop in the $m=2$ Fourier amplitude of the disk surface
  density.  The bars are not destroyed by the buckling but continue to grow
  until the present day.  Bars are largely absent when the disk mass is
  reduced by a factor of two or more; the relative disk-to-halo mass is
  therefore a primary factor in bar formation and evolution.  A
  subset of the disks is warped at the outskirts and contains prominent
  non-coplanar material with a ring-like structure. Many disks reorient
  by large angles between $z=1$ and $z=0$, following a coherent
  reorientation of their inner haloes. Larger reorientations produce
  more strongly warped disks, suggesting a tight link between the two
  phenomena. The origins of bars and warps appear
  independent: some disks with strong bars show little disturbances at the
  outskirts, while the disks with the weakest bars show severe warps.
\end{abstract}

\section{Introduction}\label{sec:intro}

At low redshift stellar disks are ubiquitous in galaxies of similar mass to 
the Milky Way. The properties of these disks show many regularities (see 
the review by \citealt{kruitfreeman11} and references therein), but 
reproducing these regularities in simulations of galaxy formation in a 
$\Lambda$CDM universe has been a major challenge, although recent improvements 
in subgrid physics and numerical resolution have begun to result in disks 
that resemble those seen in dwarf galaxies and the Milky Way (e.g.,
\citealt{governato10, agertz11, guedes11}).  Each disk is simulated under a 
specific schematic representation of the feedback physics, and at considerable
computational costs.

Results from the Aquila comparison project \citep{scannapieco12}, however,
suggest that the latest hydrodynamic simulations can not yet uniquely
predict the properties of baryons in a galaxy, even when the assembly
history of its host halo is fully specified.  When a common dark matter
halo is resimulated with gas with nine cosmological hydrodynamic codes,
large code-to-code differences are found in the $z=0$ stellar mass, size,
morphology, and gas content.  The variations are mainly due to the
different implementations of feedback.  The feedback models that are more
effective at suppressing the baryonic mass in the galaxy are better at matching
the observed scaling relations of the global properties of the galaxies, but
they do not necessarily produce more realistic disks.

The goal of this paper is to investigate the gravitational interactions
between live stellar disks and their dark matter haloes, when these are
themselves evolving in their full cosmological context. Rather than
focusing on how to form a disk with realistic structure, we choose to
study how preformed stellar disks are influenced by the cosmologically
driven evolution of the haloes in which they are embedded. We use
haloes from the Aquarius Project (Springel et al. 2008) because they
provide a particularly well studied set of high resolution haloes of
similar mass.

We simulate each disk-halo system in two phases, first by allowing each
Aquarius dark matter halo to respond adiabatically to a growing rigid disk
potential between $z=1.3$ and 1.0, and then by inserting a live disk of
stellar particles that are evolved with the dark matter particles from
$z=1.0$ to today.  We insert disks in four Aquarius haloes (A, B, C, and
D), with the disk normal initially aligned with the minor or major axis 
of the inner halo as measured at $z=1.3$.  The main results in this paper 
are drawn from these eight simulations; several additional runs are also 
performed to test numerical convergences and to assess the impact of different 
disk models.

Our study complements several earlier papers that used the Aquarius haloes
for numerical explorations of the formation and evolution of galaxies like
our own.  \citet{scannapieco09} resimulated a number of the original
Aquarius haloes including the effects of a realistic multi-phase,
star-forming gas.  Disks were found to form at $z \ga 2$, but were often
destroyed again by mergers or instabilities, depending on the particular 
assembly history of each object.  \citet{tissera10} analyzed the effects of the
baryons on the dark matter halos in the same set of simulations.  They
found the response of the haloes to baryon condensation to depend on the
specifics of how the halo was constructed.  \citet{cooper10} tagged subsets
of dark matter particles in the Aquarius haloes with stellar populations
assigned according to a semi-analytic model of galaxy formation and studied the
spatial and kinematic properties of the stellar haloes.
\citet{aumer12arxiv} investigated properties of stellar disks which they formed 
in the Aquarius haloes by allowing a smooth, spherically symmetric cloud of 
rotating gas to cool and condense at halo center.

Numerous earlier studies have inserted live stellar disks into dark matter
haloes, most of them with an idealized setup and/or environment for the
disks and haloes.  Some $N$-body studies focused on isolated disk and halo
systems and studied internal or secular evolution, e.g., \citet{dubinski94,
  athanassoula02, debattista06, berentzen06, machado10}.  Other studies
presented results for the effects of external perturbations such as
infalling satellites on an existing live disk in a halo, e.g.,
\citet{quinn86, mihos95, walker96, sellwood98, velazquez99, font01,
  benson04, gauthier06, curir06, curir07, bournaud07, hopkins08,
  read08, kazantzidis08, kazantzidis09, purcell09, chakrabarti09,
  moster12}.  Our study takes advantage of some of the most realistic and
well resolved dark matter haloes currently available.  This setup allows us
to investigate the full dynamical interactions of live stellar disks and
live haloes in the presence of cosmological mergers and large-scale tidal
fields, without having to make assumptions about the triaxiality and
equilibrium state of the halo, or about the orbits and masses of the
infalling satellites.

The organization of this paper is as follows.  Section~\ref{sec:methods}
describes the methodology used to insert the stellar disk into the halo and
the parameters used in the simulations.  The effects of the growing rigid
disk on the axial lengths and triaxiality of the haloes between $z=1.3$ and
1.0 are described in Section~\ref{sec:rampeffects}.  Results for the
evolution of the live stellar disks in our main set of simulations are
presented in Section~\ref{sec:barsanddisks}.  Here we investigate the
radial and vertical structures of the disks (Sec.~4.1), the formation and
evolution of stellar bars (Sec.~4.2), the velocity profiles and heating due
to the bars and other transient features in the disk (Sec.~4.3), and the
warps in the outskirts of the disks and the correlation of the warps with
the large-scale rotation of the haloes (Sec.~4.4).  Results from
alternative disk models such as disks with lowered stellar masses and disks with
non-cooling gas are given in Section~\ref{sec:altdisks}.

\begin{table*}
  \label{table:simparm}
  \begin{tabular}{lccccccc}
   \hline
   Run Name & Halo & Disk & $M_d$ & $R_d$ & $z_d$ & $Q_{bar}$ & Notes\\
   & & Orientation & [$10^{10} \msun$] & [kpc] & [kpc] & [$z=1.0$] & \\
   \hline
   AMinor  & A5 & Minor & 5.00 & 3.00 & 0.60 & 0.98 &\\
   AMajor  & A5 & Major & 5.00 & 3.00 & 0.60 & 0.99 &\\
   AMinorHalf  & A5 & Minor & 2.50 & 2.38 & 0.48 & 1.13 &\\
   AMinorThird  & A5 & Minor & 1.67 & 2.08 & 0.42 & 1.23 &\\
   AMinorGas  & A5 & Minor & 5.00 & 3.00 & 0.60 & - & initial gas fraction of $0.4$\\
   AMinorHR & A5 & Minor & 5.00 & 3.00 & 0.60 & 0.98 & $5 \times 10^5$ particles in disk\\
   AMinor09 & A5 & Minor & 5.00 & 3.00 & 0.60 & 1.00 & Transition to live disk at $z = 0.9$\\
   ANoDisk & A5 & - & - & - & - & - & No disk\\
   \hline
   BMinor  & B5 & Minor & 5.00 & 3.00 & 0.60 & 0.83 &\\
   BMajor  & B5 & Major & 5.00 & 3.00 & 0.60 & 0.81 &\\
   BNoDisk & B5 & - & - & - & - & - & No disk\\
   \hline
   CMinor  & C5 & Minor & 5.00 & 3.00 & 0.60 & 1.03 &\\
   CMajor  & C5 & Major & 5.00 & 3.00 & 0.60 & 1.03 &\\
   CMinorHalf  & C5 & Minor & 2.50 & 2.38 & 0.48 & 1.15 &\\
   CMinorThird  & C5 & Minor & 1.67 & 2.08 & 0.42 & 1.26 &\\
   CMinorGas  & C5 & Minor & 5.00 & 3.00 & 0.60 & - & initial gas fraction of $0.4$\\
   CNoDisk & C5 & - & - & - & - & - & No disk \\
   \hline
   DMinor  & D5 & Minor & 5.00 & 3.00 & 0.60 & 0.93 & \\
   DMajor  & D5 & Major & 5.00 & 3.00 & 0.60 & 0.92 & \\
   DNoDisk & D5 & - & - & - & - & - & No disk\\
   \hline
  \end{tabular}

  \caption{Simulation Parameters:
    ``Halo'' refers to the Aquarius dark matter halo (and resolution level) 
    used in the simulation.
    ``Disk Orientation'' indicated whether disk normal is initially aligned with 
    the major or with the minor axis of the halo at $z=1.3$.
    Parameters $M_d$, $R_d$, and $z_d$ are the total mass, scale length, and scale
    height of the disk at the start of the live phase at $z=1.0$, respectively. 
       The bar criterion parameter $Q_{bar}$ 
    is measured at $z=1.0$; it is defined in eq.~(5) and $Q_{bar}> 1.1$ 
    indicates stability against bar formation.}
\end{table*}

\section{Methods and Simulation Parameters}\label{sec:methods}

To study the evolution of stellar disks in a full cosmological context, we
begin with a suite of dark matter haloes from the Aquarius simulations
\citep{springel08}.  We then add a stellar disk to each Aquarius halo in
two phases.  During the first phase -- starting at $z = 1.3$ and ending at
$z = 1.0$ -- the dark matter is allowed to react to a rigid disk potential
whose mass increases from zero at $z=1.3$ to the desired final mass at
$z=1.0$.  At the start of the second `live' phase at $z = 1.0$, the rigid
disk is replaced with live simulation particles.  This live disk is then
evolved self-consistently along with all the dark matter particles to
$z=0$.  Below we describe each step in detail.

\subsection{Dark Matter Haloes}

The high-resolution zoom-in simulations of dark matter haloes in the
Aquarius Project are chosen from a lower resolution version of the
Millennium-II Simulation \citep{boylankolchin09}.   These resimulated
  haloes are chosen randomly from those with a mass similar to that of the
  Milky Way, and which does not have a massive close neighbor at $z =
  0$. These haloes are therefore good candidates to host Milky Way-like
  disk galaxies.  The cosmological model is $\Lambda$CDM with
$\Omega_{\Lambda} = 0.75$, $\Omega_m = 0.25$, $\Omega_b = 0.04$, $\sigma_8
= 0.9$ and $H_0 = 73 \kms \rm{Mpc}^{-1}$.

In this work we select haloes A, B, C and D at resolution level 5.
This resolution level corresponds to dark matter particle masses of
$\approx 3 \times 10^6 \msun$, and a comoving gravitational force softening
length of 685 pc.  The virial mass (defined using an overdensity 
of $200 \rho_{crit}$) of the four dark matter haloes at $z=0$
is $1.49\times 10^{12} M_\odot, 7.11\times 10^{11} M_\odot, 1.61\times
10^{12} M_\odot$, and $1.49\times 10^{12} M_\odot$ for A, B, C and D,
respectively.  Despite their comparable final masses, the four haloes have
different merger histories.  While each halo has accreted a fair number of
small subhaloes during the course of the simulation, the level of activity
for infalling satellites above one tenth the disk mass $M_d$ (chosen to be
$M_d=5\times 10^{10} M_\odot$ for our fiducial disk model) varies from halo
to halo.  Haloes A and B are relatively quiet after $z \sim 0.7$ when a
subhalo of mass $\sim 0.3 M_d$ impacts in each.  Halo C has a number of
larger subhaloes in the latter part of the simulation ($z < 0.5$), where
the two most massive subhaloes have masses of $0.22 M_d$ and $0.16 M_d$.
Halo D has the most active history, with appreciable encounters spread
throughout the simulation.  Although these impacts are defined at the halo 
level, and thus are not necessarily close enough to the disk to have a 
dynamical effect, they are candidates for study in future work linking 
specific halo substructures to features of the disk evolution.

\subsection{Phase 1: Adding a Rigid Disk}

We modified a version of GADGET-2 \citep{springel05} to add a rigid stellar
disk potential to the existing dark matter particles.  The potential
corresponds to an exponential disk with density:
\begin{equation}
\label{eqn:diskmodel}
\rho(R,z,t) = \frac{M_d(t)}{4 \pi R_d^2 z_d} e^{-R/R_d}\, 
     {\rm sech}^2 \left( \frac{z}{z_d} \right)  \,,
\end{equation}
where $R_d$ and $z_d$ are the scale length and scale height of the disk,
respectively, and $M_d(t)$ is the mass of the disk at time $t$.  The mass
of the disk is increased linearly in the scale factor, which is
approximately linear in time, from zero at $z=1.3$ to the total mass,
$M_d$, at $z=1.0$.

The center of the disk is initially placed at the minimum of the
gravitational potential of the main Aquarius halo at $z = 1.3$.  The
initial velocity of the disk is set to the velocity of the potential
minimum.  To orient the disk, we first determine the principal axes of the
gravitational potential of the dark matter halo by constructing
equipotential surfaces of the dark matter particles and fitting ellipsoids
to them.  The disk normal is chosen to be aligned with either the minor or
major axis of the halo at $z=1.3$.  We refer to these two disk orientations
as ``Minor'' and ``Major'' in the rest of the paper.

During this phase of the simulation, each dark matter particle feels a force
from this rigid potential in addition to the normal gravitational
interactions with all the dark matter particles.  These extra forces are
applied in a self-consistent manner, with the disk center experiencing the
third-law force pairs from each dark matter particle.  This procedure
assures that linear momentum is conserved and the disk center moves with
the center of the main halo.

The orientation of the rigid potential is fixed during this phase.  This
choice appears appropriate since the dark matter haloes do not rotate much
between $z=1.3$ and 1.0.  We do find, however, that the principal axes of
the haloes can drift by up to 20 degrees between $z = 1.3$ and 1.0.  This
drift implies that if our disk's normal is initially aligned with one of
the halo's axes at $z=1.3$, it may become slightly misaligned by $z=1.0$.
We have performed a test run in which the live disk's normal is exactly
aligned with the minor axis of halo A at $z=1.0$ (rather than at $z=1.3$ as
in the fiducial runs).  We find this slight adjustment in the initial
alignment to make little difference in the subsequent evolution of the disk.

\subsection{Phase 2: Live Disk}

Phase 1 above allows the dark matter particles to respond adiabatically to
the presence of a growing rigid disk potential between $z=1.3$ and 1.0.
Phase 2 begins at $z=1.0$ when we add live disk particles to the
simulations.

The initial conditions of the live disk particles are generated by creating
a model of the full potential of the system, and finding an approximate
solution to the Jeans equations.  The potential of the disk is easily
computed from the model density, but for the halo potential we perform a
fit to the dark matter particle potentials at the end of phase 1 at $z=1.0$
(excluding the contribution to the potential from the rigid disk).  The
halo particles are projected into coordinates centered on the disk with the
$z$ axis aligned with the disk normal.  The potential values of the halo
particles are fit to the following functional form:
\begin{equation}\label{eqn:halopot}
    \Phi = v_c^2 \ln{\left[1 + \frac{u^2}{R_c^2}\right]} \,,\qquad
    u^2 = z^2 + \frac{x^2 + y^2}{a_3^2} \,,
\end{equation}
where $u$ is an ellipsoidal coordinate,
and $v_c$, $R_c$ and $a_3$ are parameters of the fit.  This definition of
$u$ forces the description of the background halo potential to have one of
its principal axes along the disk normal.  This is a good assumption for
the evolved halo in both the major and minor orientation.  This definition
also assumes a potential which is axisymmetric with respect to the disk normal
direction.

Once the live disk particles are added at $z=1.0$, the rigid disk potential
is turned off and both the stellar and dark matter particles are allowed to
evolve self-consistently until $z=0$.

To test the stability of the live disks generated with this method, we have
evolved the disk particles in isolation subject to a static background
potential equal to the fitted halo potential at $z=1.0$ in
equation~(\ref{eqn:halopot}).  The evolved disk is stable over a long
time and develops no significant structure.  The strong bar formation in
most of the live simulations that we report below must therefore depend on
reinforcement by a coherent response of the halo to the growing
perturbation.

\subsection{Disk Parameters}\label{sec:simparm}

We have run a suite of simulations with a range of disk parameters and
properties to assess the robustness of our results.  The main parameters of
the simulations are summarized in Table~\ref{table:simparm}.

\begin{figure*}
\includegraphics[width=170mm]{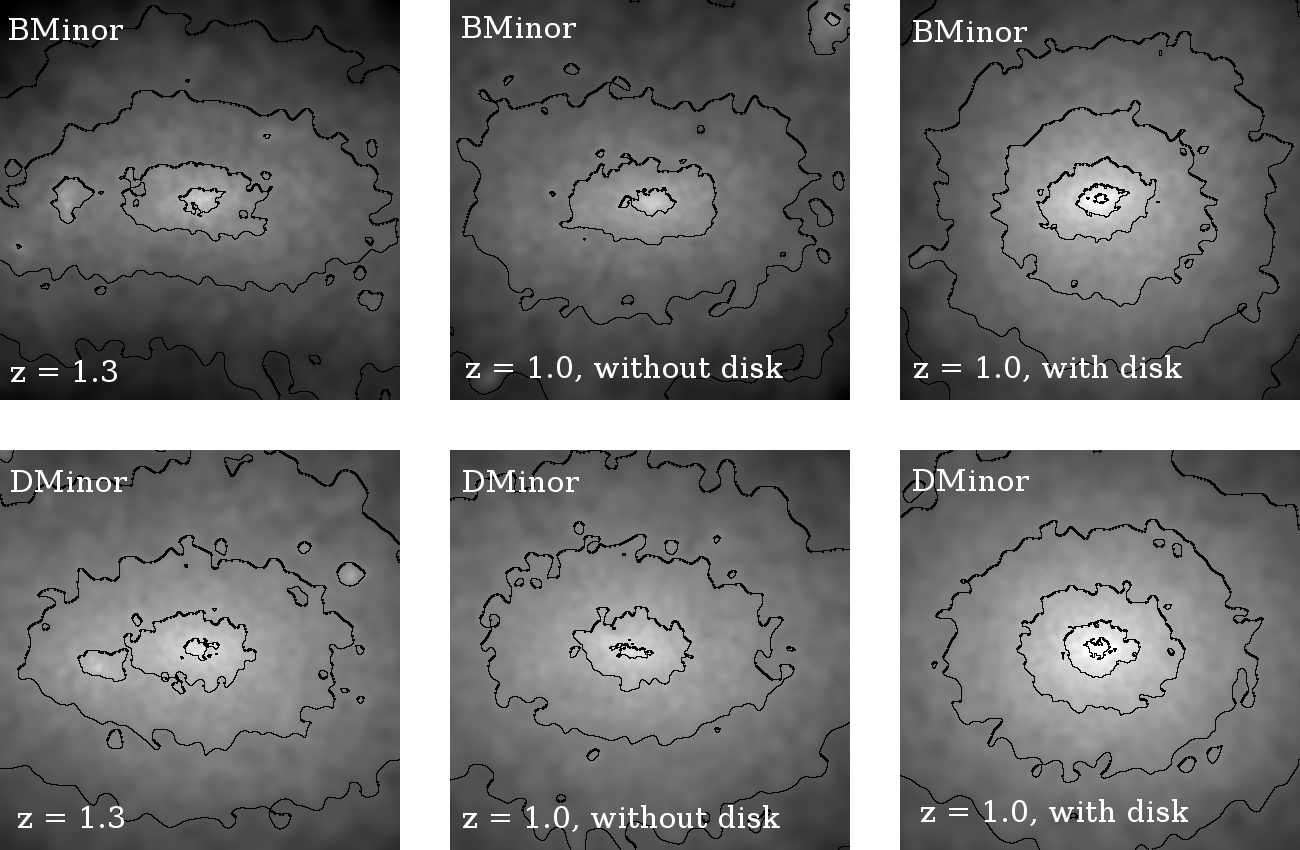}
\caption{Surface density of the dark matter particles in halo B (top) and
  halo D (bottom), projected along their minor axes.  The images are 40 kpc
  across.  In the pure dark matter run without a disk (left and middle
  columns), the haloes are prolate (in the inner 40 kpc) and the shapes
  evolve little from $z=1.3$ to 1.0.  However, when a rigid disk is
  introduced (with disk normal aligned with the minor axis) adiabatically
  between $z=1.3$ and 1.0 (right column), the dark matter haloes are
  symmetrized in the disk plane and become oblate by $z=1.0$.  The inner
  haloes also become denser as illustrated by the more closely spaced
  constant-density surfaces in the right column.}
\label{fig:HaloImages}
\end{figure*}

Our fiducial disk model assumes a stellar disk mass of $M_d = 5 \times
10^{10} \msun$, a scale length of $R_d = 3$ kpc, and a scale height of $z_d
= 0.6$ kpc.  These values are chosen to resemble the Milky Way's disk today
\citep{juric08, mcmillan11}.  This should also be a reasonable choice for
$z =1.0$ as disk size does not exhibit strong evolution between $z \sim
1.0$ and 0 \citep{trujillo06b}.  Each realization of the disk uses $2
\times 10^5$ stellar particles with a comoving gravitational force
softening of 137 pc for these particles.
For each halo, we run two separate simulations for two disk orientations,
in which the initial disk normal is aligned either with the minor or with
the major axis of the halo at $z=1.3$.

For a direct comparison between simulations without and with a disk, we run
dark-matter-only simulations with the particle positions and velocities
from the Aquarius $z=1.0$ snapshot as initial conditions until $z=0.0$
(labeled with `NoDisk' in Table~\ref{table:simparm}).  Although the
particle positions and velocities at $z=0.0$ are not identical to those in
the original Aquarius runs due to differences in machine and numerical
details, the properties of the haloes are very similar.

To quantify the dependence of our results on the assumed disk mass, we run
additional simulations with one half and one third the fiducial disk mass
in the minor orientation in the A and C halos (labeled `Half' and `Third').
The scale lengths are reduced in accordance with the observed relation, $R_d
\propto M_d^{1/3}$ \citep{shen03}.

Two additional simulations are performed with haloes A and C that include a
gaseous component with initial gas fraction of $f_g = 0.4$.  This component
is comprised of $10^5$ SPH particles. The gas particles are initialized
with the same radial profile as the stellar component, but the vertical
structure is initialized in hydrostatic equilibrium.  For simplicity, the
gas used in these runs is not star forming and does not cool.  We include
this component to test if a gas component helps stabilize the stars in a
disk.  A more realistic simulation would turn most of the gas at $z=1$ into
stars by $z=0$.

 To determine the effect of changing the time at which the disk becomes
  live, we have run a simulation in which the rigid disk potential is
  transitioned to a live disk at $z = 0.9$ rather than $z=1.0$ (`AMinor09'
  in Table~\ref{table:simparm}).  In this run the rigid disk potential
  grows from zero to its final mass over the period of $z = 1.3$ and 0.9.
  Other paramaters are identical to those used in the AMinor simulation.

Finally, to assess if our results are converged with respect to the stellar
disk particle number, we perform a simulation of the AMinor disk with
$5\times 10^5$ instead of $2\times 10^5$ stellar particles.  This run is
listed as `AMinorHR' in Table~\ref{table:simparm}.

\begin{figure*}
\includegraphics[width=180mm]{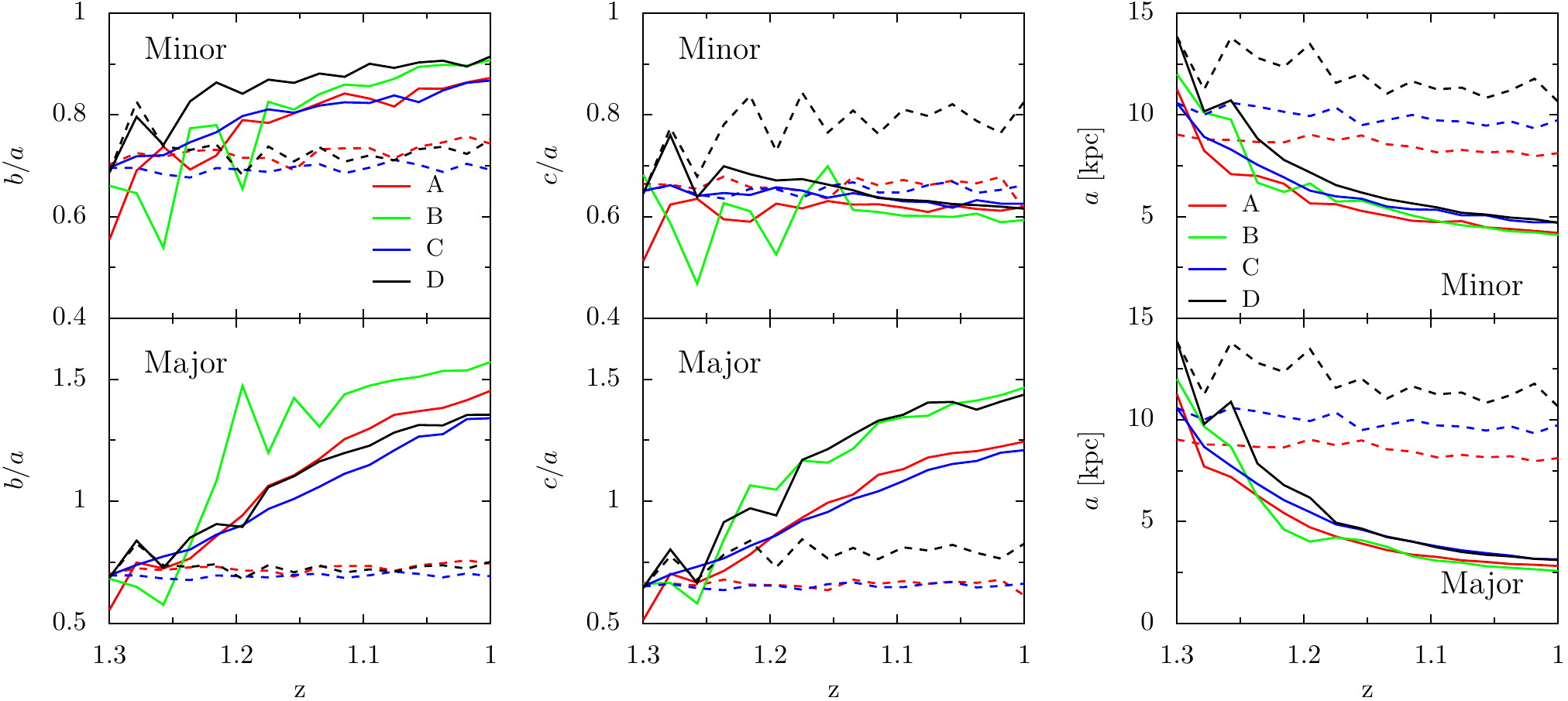}
\caption{Effects of the growing rigid disks on the shapes of dark matter
  haloes between $z=1.3$ and 1.0.  The left and middle columns show the
  time evolution of the axial ratios $b/a$ and $c/a$, respectively; the
  right column shows the effect of the disks on the initial major axis $a$
  (for a given equipotential surface) of the haloes.  Within each panel,
  the dashed curves show the axes for the original Aquarius haloes A (red),
  B (green), C (blue), and D (black) between $z=1.3$ and 1.0.  The ratios
  stay nearly constant with $c \sim b \sim 0.7 a$.  The solid curves show
  how the insertion of a growing rigid disk between $z=1.3$ and 1.0 shrinks
  all three axes of the dark matter haloes and modifies the axial ratios.
  The modifications depend on whether the disk normal is initially aligned with
  the minor (top panels) or with the major (bottom panels) axis of the halo at
  $z=1.3$.  Since the axial lengths of the haloes change with time, the
  initial major and minor axes may not stay as the major and minor axes at
  all times.  For ease of comparison, we name the principal axes $a, b, c$
  (with $a\ge b \ge c$) at $z=1.3$ and plot the subsequent evolution of the
  {\it same} axes.  The ratios $b/a$ and $c/a$ can therefore become greater
  than 1 at later times.  The axial ratios for the original halo B are very
  noisy and are not plotted because a nearby subhalo twists its major axis
  as it falls into the halo.}
\label{fig:proandflat}
\end{figure*}

\section{Effects of Rigid Stellar Disks on Dark Matter Haloes}
\label{sec:rampeffects}

In this section we discuss the effects of the rigid disk potential on the
shapes and orientations of the dark matter haloes during the disk growing
phase of the simulations ($z = 1.3$ to 1.0).  By comparing our simulations
with the original Aquarius dark-matter-only haloes, we can quantify how the
presence of stellar disks in more realistic simulations alters the
distributions of the dark matter.

To measure the triaxiality of each halo, we determine the principal axes
by constructing equipotential surfaces (excluding the contribution to the
potential from the rigid disk) near the center of the halo with the
procedure discussed in Sec~2.3.  In general, the equipotential surfaces are
fit well by ellipsoids, failing to be ellipsoidal only in the presence of
significant substructure.  In haloes B and D, there are significant
subhaloes near the position of the disk initially, so the fit to an
ellipsoid fails for a few potential bins.  Nevertheless, the fit performs
well for most of the potential bins (and hence most radii) and for most
times.

Figs.~\ref{fig:HaloImages} and \ref{fig:proandflat} illustrate the effects
of the rigid disk on the triaxiality of the dark matter haloes.
Fig.~\ref{fig:HaloImages} shows the surface density of dark matter
particles in the inner 40 kpc of haloes B (top) and D (bottom) at the
initial time $z=1.3$ (left column), and at $z=1.0$ for runs without the
disk (middle column) and with the disk (right column).
Fig.~\ref{fig:proandflat} plots the time evolution of the axial lengths of
each halo, where the orientations as well as axial lengths are computed at
each snapshot.  The left two columns of Fig.~\ref{fig:proandflat} show the
axial ratios determined from equipotential surfaces at $\sim 5$ kpc, and
the right column shows the evolution of the length of the initial major
axis for a chosen equipotential surface.  The results are representative of
the other equipotential surfaces.

Without stellar disks, Figs.~\ref{fig:HaloImages} (left two panels) and
\ref{fig:proandflat} (dashed curves) show that the inner $\sim 50$ kpc of
all four dark matter haloes from the Aquarius simulations are prolate with
axial ratios of $c \sim b \sim 0.7 a$.  The ratios stay quite constant
between $z=1.3$ and 1.0.

When a rigid disk is introduced, all three axes of the halo shorten in
response to the added mass.  This contraction is illustrated by the more
closely spaced equidensity contours in the right column of
Fig.~\ref{fig:HaloImages}, and by the shrinking length of the initial major
axis in the right column of Fig.~\ref{fig:proandflat}.

Despite the overall contraction of the inner halo in response to the disk,
how the axial ratio of the halo is modified depends on the orientation of
the disk.  When the disk normal is initially aligned with the minor axis of
the halo, we find the presence of the disk to nearly symmetrize the axial
ratio of the dark matter halo in the disk plane, as shown by the rounder
contours in the right column of Fig.~\ref{fig:HaloImages}.  The solid
curves in the upper panels of Fig.~\ref{fig:proandflat} show the
corresponding evolution in the axial ratios: $b/a$ evolves from $\sim 0.7$
towards 0.9, while $c/a$ remains relatively constant at $\sim 0.6$ to 0.7
between $z=1.3$ and 1.0.  The inner $\sim 50$ kpc of the haloes therefore
changes from being prolate to nearly oblate in the presence of a disk in the
minor orientation.

\begin{figure*}
\includegraphics[width=170mm]{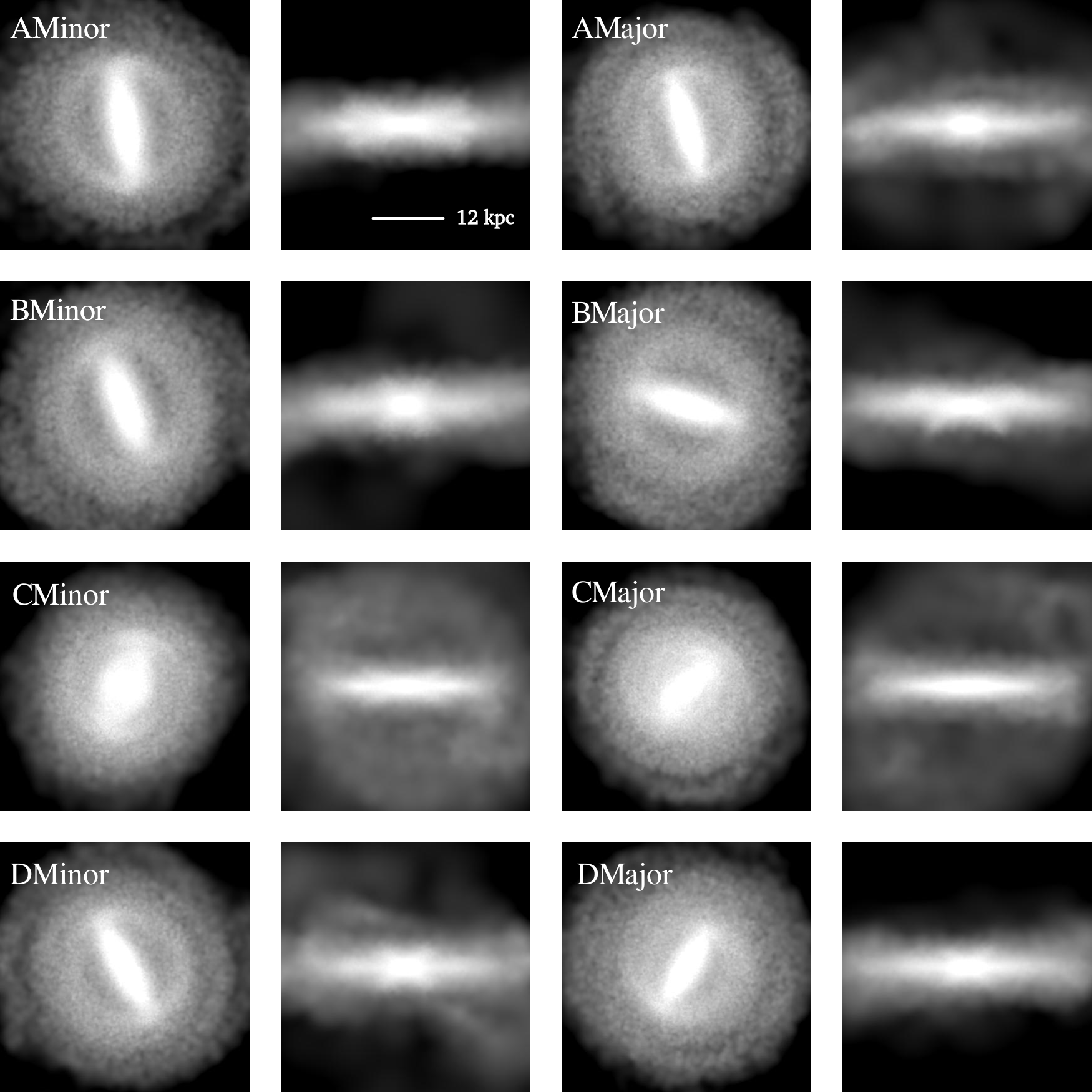}
\caption{Surface density of the $z=0$ stellar disk, viewed faced-on
  (columns 1 and 3) and edge-on (columns 2 and 4), for haloes A, B, C and D
  (from top down).  Each image is 43 kpc on a side. ``Minor'' and ``Major''
  indicate whether the disk normal is initially aligned with the minor or
  major axis of the dark matter halo at $z=1.3$.}
\label{fig:Images}
\end{figure*}

When the disk normal is initially aligned with the major axis of the halo,
the halo in the disk plane starts out quite symmetric since $b\sim c$ at
$z=1.3$.  As the solid curves in Fig.~\ref{fig:proandflat} show, the
initial major axis $a$ contracts much more in this orientation than in the
minor orientation (right column) in the presence of the disk, increasing
the axial ratios $b/a$ and $c/a$ but keeping the ratio of $b/c$ relatively
constant (left two columns).  The haloes become nearly spherical at $z\sim
1.2$, after which the initial major axis becomes the minor axis as $b/a$
and $c/a$ both exceed unity.  Note that for ease of comparison, we name the
principal axes $a, b, c$ (with $a \ge b \ge c$) at $z = 1.3$ and plot the
subsequent evolution of the {\it same} axes.  The haloes therefore
evolve from being prolate to oblate in the presence of a disk in the major
orientation.  The difference here is the disk normal that was initially
aligned with the halo's major axis at $z=1.3$ ends up being aligned with
the halo's minor axis at $z=1$, and this occurs because of changes in the
axial lengths of the halo rather than re-orientation of the disk.

Our results for the shapes of haloes without and with a stellar disk are in
broad agreement with previous work.  Dark matter haloes in $N$-body
simulations generally have a range of shapes with a preference for
prolateness over oblateness; the shapes can also depend on the radius,
varying from prolate in the inner regions to triaxial or oblate in the
outer parts (e.g., \citealt{frenk88, dubinski91, warren92, colelacey96,
  jingsuto02, bailin05, allgood06, hayashi07, bett07, veraciro11}).  The
presence of a disk in simulations has been shown to wash out the
prolateness of the halo and make the inner haloes either more spherical or
oblate (e.g., \citealt{dubinski94, berentzen06, debattista08,
  kazantzidis10, tissera10, abadi10}).  An analysis of the characteristic
frequencies of orbits in dark matter haloes \citealt{valluri10} suggests
that the change in halo triaxiality is due primarily to changes in the
shapes of individual orbits of collisionless particles in response to a central
baryonic component.

In addition to changes in the axial ratios, adding a stellar disk can also
reorient the halo's equipotential surfaces between $z=1.3$ and 1.0.
Comparing the haloes in our simulations with disks and those in the
original Aquarius simulations, we find that with the exception of halo B,
the major axis of the halo does not experience significant reorientation in
the presence of a growing disk.  In halo B, a subhalo is visible at
$\sim 15$ kpc from the disk center at $z=1.3$ in
Fig.~\ref{fig:HaloImages}.  In the original Aquarius halo, this subhalo
twists the major axis of halo B as it falls into the halo.  The axial
length for halo B is therefore quite noisy and is not plotted in
Fig.~\ref{fig:proandflat}.  When a stellar disk is added, we find the
major axis to be more stable due to the influence of the rigid disk.

\section{Live Stellar Disks, Bars, and Warps}
\label{sec:barsanddisks}

In this section we present the results for the evolution of the live
stellar disks in the Aquarius dark matter haloes from $z=1.0$ to 0.0.

To determine the disk plane at each output, we bin the disk particles
in radius, and draw an imaginary plane through the center of mass of
the entire disk.  The sum of the squared distances of each particle in the
given bin to the plane is computed, and the normal to the plane is varied
to minimize this sum.  In this way, we find a disk normal for each radial
bin of the disk.  This process is performed for each output and a history
of the disk's orientation is constructed.  As reported below, we find this
procedure to produce disk normals that are nearly identical in the central
region of the disk, but the outskirts can be warped and have a very different
orientation.   We have also computed the angular momentum vector for 
each disk using all the disk particles.  This vector is aligned to within
2 degrees with the normal of the inner disk.

\subsection{Structural Properties of Live Disks}

\begin{figure*}
\includegraphics[width=170mm]{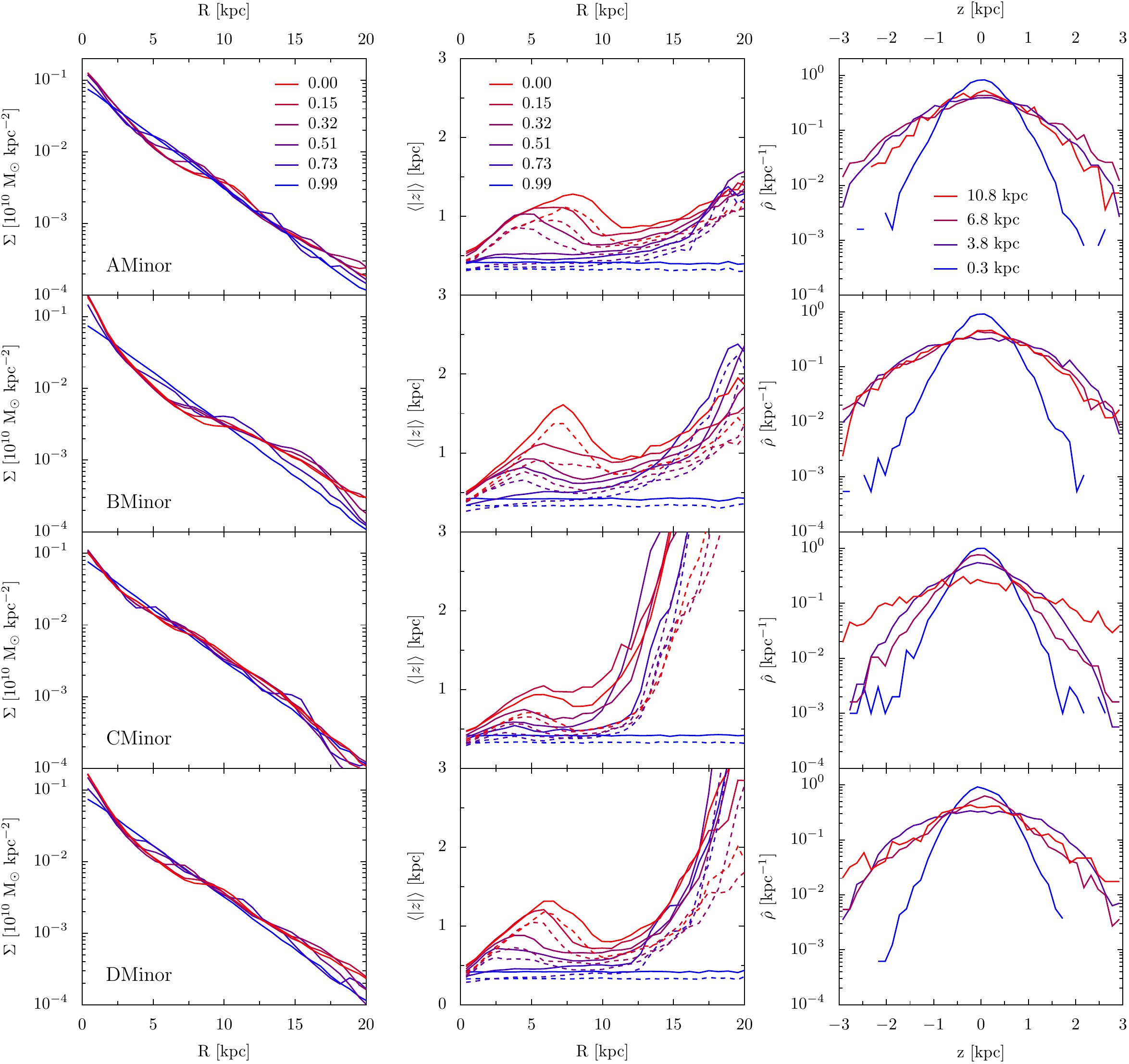}
\caption{Structural properties of the disks in the minor orientation for
  haloes A, B, C, and D (top down).  The projected surface density of the
  disk as a function of radius (left column) is quite stable between
  $z=0.99$ (blue) and 0.0 (red), with a steepening in the inner few kpc due
  to the stellar bars.  The middle column shows two
  measures of the height of the disk, the mean (solid) and median (dashed)
  distance of the disk particles from the disk plane, as a function of
  radius for six redshifts during the simulations.
  The vertical profiles of the disk at $z=0$ (right
  column) become broader with increasing radii.  }
\label{fig:MinorStructure}
\end{figure*}

\begin{figure*}
\includegraphics[width=170mm]{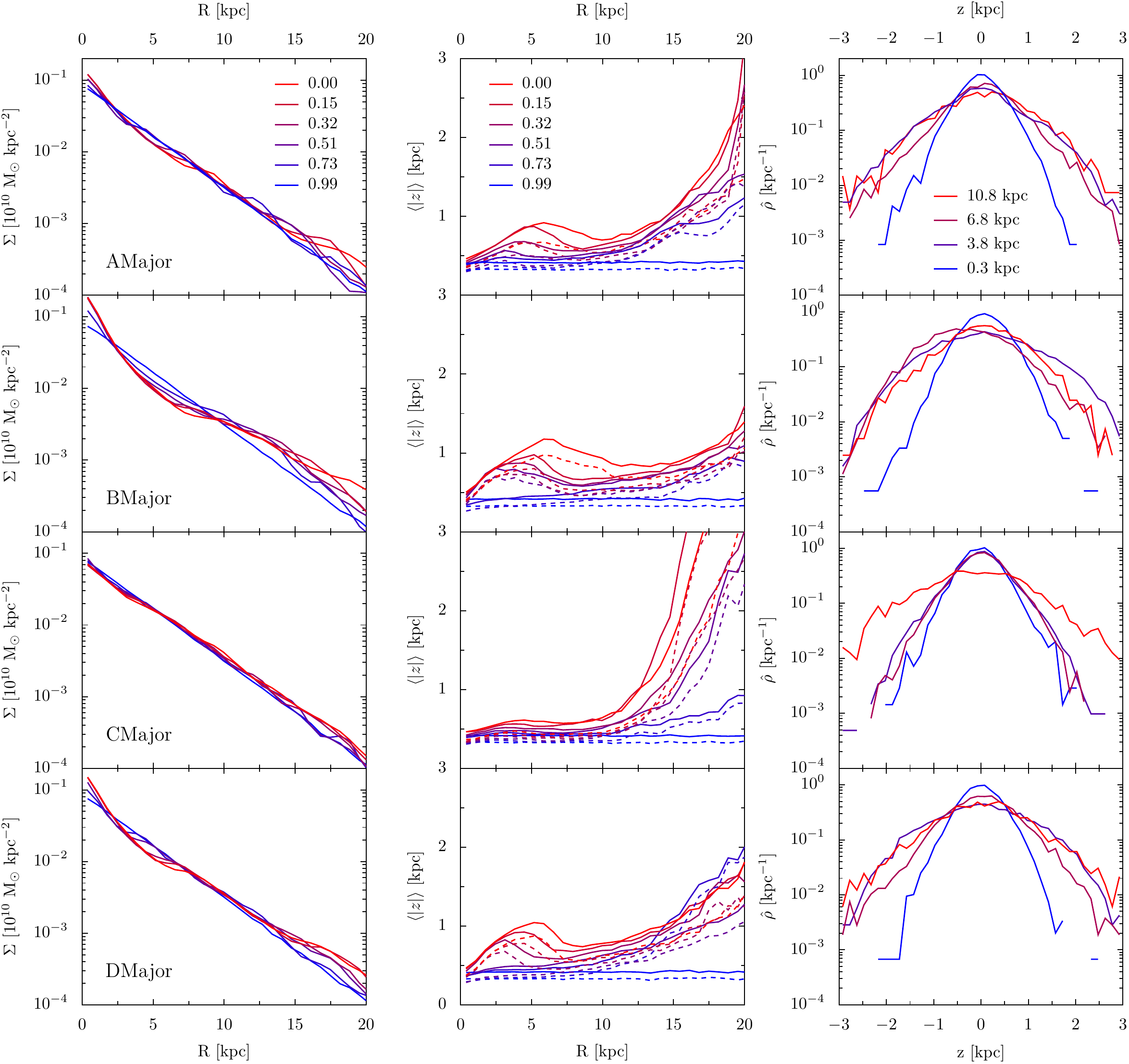}
\caption{Same as Fig.~\ref{fig:MinorStructure} but for the disks whose normal is initally 
  aligned with the major axis of the halo at $z=1.3$.
}
\label{fig:MajorStructure}
\end{figure*}

Images of the final ($z=0$) surface densities of the live disks are plotted
in Fig.~\ref{fig:Images}.  For completeness, we show both the face-on and
edge-on views of each disk in the four haloes.  The left two columns show
the disks that are initially aligned with the minor axis of the halo; the
right two columns show the major axis runs.  The images (43 kpc on a side)
clearly show that a bar extending to at least $2 R_d$ has formed in each
disk, with the C disks containing the weakest bars, and the minor disks
containing slightly longer bars than the major disks.  We will further
quantify the bar strength and evolution in \S~4.2.

In addition to the bars, many edge-on disk galaxies in the local universe
exhibit X-shaped structures or peanut-shaped bulges (e.g.,
\citealt{whitmore88, kuijken95, bureau99, bureau06}).  These features are
also seen in Fig.~\ref{fig:Images}.  Their existence may be related to the
buckling of the bar that heats and ejects disk material
\citep{combes81,raha91, mihos95}.

The edge-on images of the disks in Fig.~\ref{fig:Images} reveal a large
variation in the amount of material out of the plane of the disk.  The
CMinor, DMinor, AMajor, and CMajor haloes contain a prominent diffuse
component of stars extending to tens of kpc outside the disk plane, whereas
this component is nearly absent in the other four haloes.  We will examine
the kinematic properties of these stars in \S~4.3 and the connection to the
outer warps of the disk plane in \S~4.4.

Fig.~\ref{fig:MinorStructure} shows snapshots of the disk surface density
(left) and mean vertical height (middle) between $z=1$ and 0, and the $z=0$
vertical profile (right) for the four ``Minor'' disks (from top down).
Fig.~\ref{fig:MajorStructure} shows the same quantities for the four
``Major'' disks.  The central surface density of the disk in the left
column is seen to increase with time when a stellar bar forms.  In
addition, there is a slight enhancement at large radii in each case.  This
effect is most pronounced in the B disks and least pronounced in the C
disks. The increase of material at large radii tends to increase the best
fit scale length $R_d$ of the disk, but we find the effect to be at most at
the $\sim 20$\% level: $R_d$ increases from the initial value of 3 kpc at
$z=1$ to 3.4, 3.7, and 3.3 kpc at $z=0$ for haloes A, B, and D,
respectively.  The scale length for halo C, by contrast, stays remarkably
constant.  We note, however, that the surface density profiles of the B and
D disks in Figs.~\ref{fig:MinorStructure} and \ref{fig:MajorStructure} at
the end of the simulations deviate significantly from the initial single
exponential form.

\begin{figure}
\includegraphics[width=80mm]{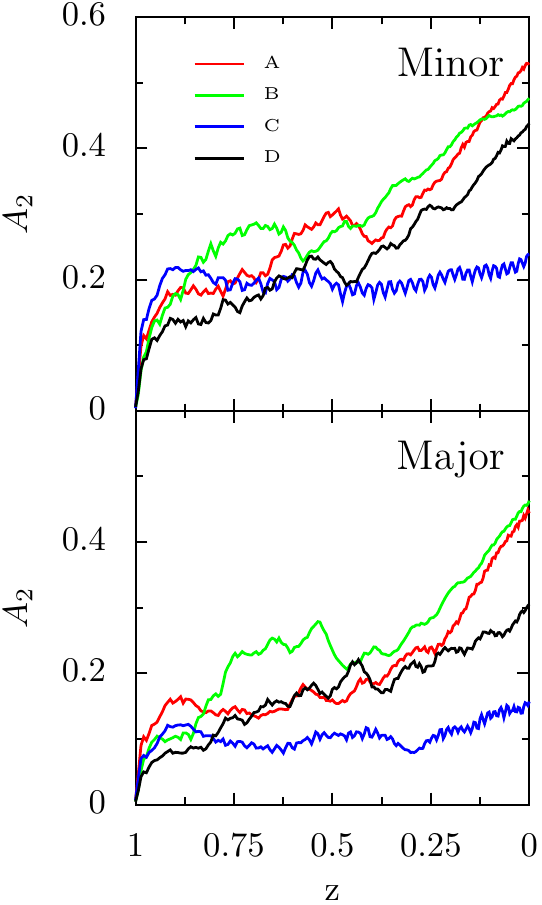}
\caption{Strengths of stellar bars measured by the $m=2$ mode, $A_2$,
  in the inner $2 R_d$ for the disks in each halo for the minor (top) and
  major (bottom) orientations.  Halo C (blue) develops the weakest bar,
  while the $m=2$ mode grows with time for haloes A (red), B (green), and D
  (black).  }
\label{fig:modesum}
\end{figure}

\begin{figure}
\includegraphics[width=80mm]{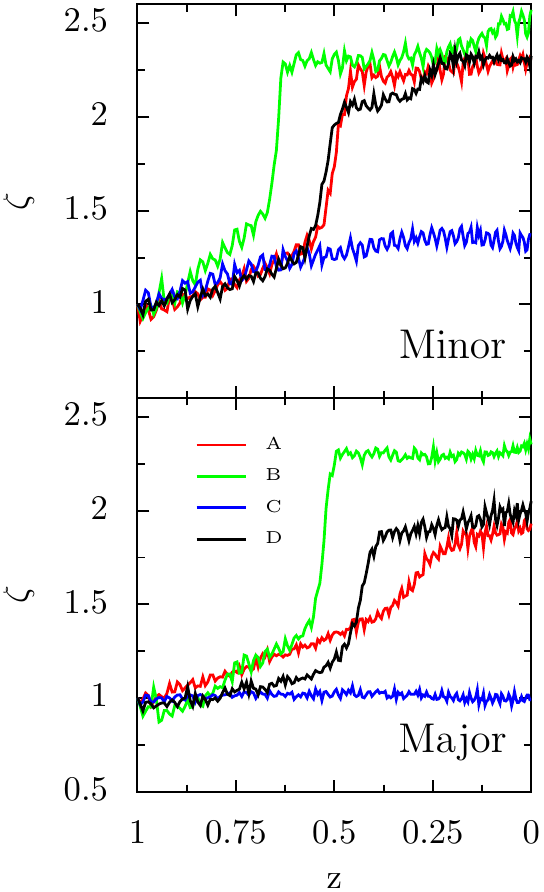}
\caption{Evolution of the vertical heating parameter $\zeta$ (defined in
  eq.~(6)) for the disks in the minor (top) and major (bottom)
  orientations.  The disks that form a strong bar, which eventually
  buckles, all show a large jump in $\zeta$ during buckling.  The two C disks
  have the weakest bars and show little heating.}
\label{fig:zeta}
\end{figure}

The middle and right columns of Figs.~\ref{fig:MinorStructure} and
\ref{fig:MajorStructure} show the vertical structures of the disks.  
  The radial profile of the mean and median vertical height $|z|$ is flat
  at the first two snapshots ($z=0.99$ and 0.73) for all eight disks, but a
  broad bump develops at $R\sim 5$ kpc in the later snapshots for disks A,
  B, D as a bar forms.  The outer profile of $|z|$ at $R\ga 10$ kpc also
  varies greatly from disk to disk.  A dramatic rise in $|z|$ is seen in
  the outer parts of CMinor, DMinor and CMajor, and to a lesser extent,
  AMajor.  This rise correlates directly with the amount of material out of
  the disk plane in Fig.~\ref{fig:Images}.  Throughout the simulation, the
  disk plane changes orientation.  While most of this reorientation is a
  coherent tumbling motion of the disk, some of the outermost material gets
  `left behind,' creating streams that are on a different plane than the
  bulk of the disk mass.  More discussion will be presented
  in \S~4.4.

  The CMajor disk is remarkable in that among the eight disks, it
  experiences the least amount of vertical thickening and heating in the
  inner 10 kpc but contains the largest amount of material out of the disk
  plane at large radii.  These results strongly suggest that the features
  at the centers of these disks and the outer material are formed from
  separate mechanisms.
  We present evidence below that the thickening at the center of the disk
  is due to a large bar that develops, whereas the material out of the
  plane is due to disk warping that results from misalignment in the inner
  and outer parts of the halo.

\subsection{Stellar Bars}

To quantify the bars that formed in the central regions of the disks, we
compute the strength of the $m=2$ Fourier amplitude of the surface density
of each disk.  The particles in the disk are binned radially in the disk
plane, and the following coefficients are computed for each bin:
\begin{eqnarray}
a_m & = & \sum_i \cos{\left(m \theta_i\right)}  \hspace{15mm} m = 0, 1, ...
     \nonumber \\
 b_m  & = & \sum_i \sin{\left(m \theta_i\right)}  \hspace{16mm} m = 1, 2, ...
\end{eqnarray}
where the sum is carried out over all the disk particles (labeled by $i$
and assumed to have equal mass) in the radial bin, and $\theta_i$ is the
azimuthal position of that particle in the disk plane.  The strength of a
given mode is $c_m = \sqrt{a_m^2 + b_m^2}$.  To quantify the strength of
the $m=2$ mode for the whole disk, we compute the relative mode strength
$A_2$ by integrating $c_2$ over the inner two scale radii:
\begin{equation}
\label{eqn:a2}
A_2 = \frac{\int_0^{2R_d} c_2(R) R dR}{\int_0^{2R_d} c_0(R) R dR}.
\end{equation}

Fig.~\ref{fig:modesum} shows the evolution of $A_2$ computed over the inner
$2 R_d$ for disks in both the minor orientation (top) and the major
orientation (bottom).  The amplitude of $A_2$ matches the expectation from
examining the images in Fig.~\ref{fig:Images}.  In particular, the bar is
weakest in the CMajor halo, with $A_2 \la 0.1$ for most of the
simulation. The CMinor disk is also distinctly stable, in which $A_2$ stays
nearly a constant at $\approx 0.2$ after a rapid initial rise from 0 to
0.2.

In contrast to halo C, the bar strengths $A_2$ for haloes A, B, and D show
a rapid initial rise from zero, followed by a $\sim 20$ to 30\% drop in
amplitude at a redshift that differs from disk to disk, and then a late
time growth leading to a final value as high as $A_2 \sim 0.5$ at $z=0$.
The dips in $A_2$ after initial bar formation have been attributed to bar
buckling instability in earlier studies such as the idealized disk and halo
systems of \citet{dubinski09} and the constrained cosmological run of
\citet{villavargas09}.  For our disks in the initial minor orientation (top
panel in Fig.~\ref{fig:modesum}), BMinor shows the earliest dip at
$z\approx 0.65$; the dips in $A_2$ for DMinor and AMinor start to occur at
$z\approx 0.5$.  For our disks in the initial major orientation, only
BMajor and DMajor exhibit noticeable dips in $A_2$ at $z\approx 0.55$ and
0.45.
  
To quantify the bar buckling instability further, we discuss in the next
subsection and show in Fig.~\ref{fig:zeta} the corresponding vertical
heating measured by the relative disk velocity dispersion in the vertical
direction in the inner $2R_d$ of the disk.  For the five disks AMinor,
BMinor, DMinor, BMajor, and DMajor, there is a clear sharp rise in disk
heating at the redshift at which $A_2$ dips in Fig.~\ref{fig:modesum}.

\begin{figure*}
\includegraphics[width=170mm]{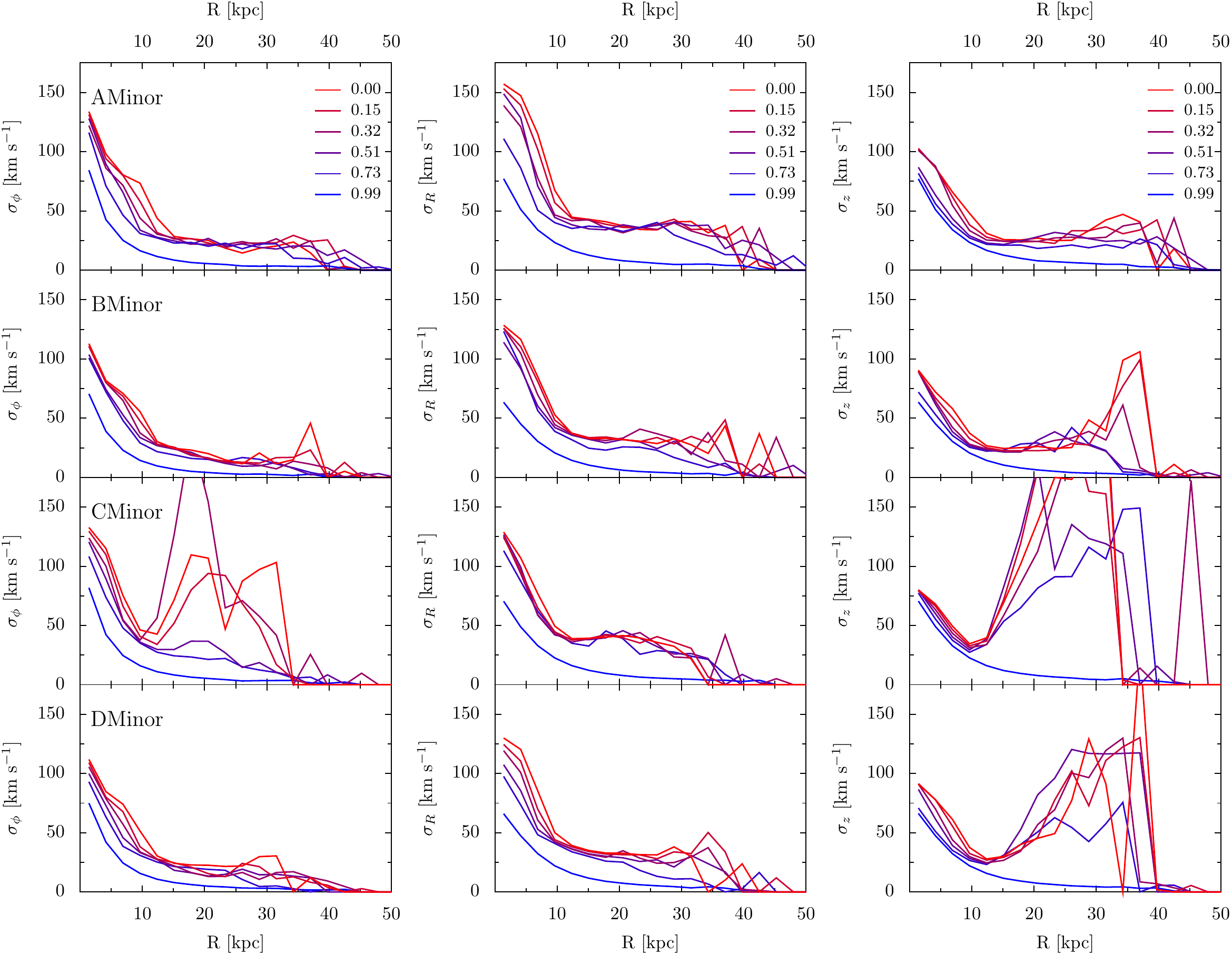}
\caption{Velocity dispersion profiles of the minor disks in haloes A, B, C and D
  (from top to bottom).  The three columns show the three velocity
  components defined by the disk: azimuthal (left), radial (middle), and
  vertical (right).  In each panel, the different curves show six outputs ranging from
  $z=0.99$ (blue) to 0.0 (red).}
\label{fig:MinorVelocity}
\end{figure*}

The formation of bars in these disks is not entirely unexpected based on
the criterion of \citet{efstathiou82}.  According to this criterion, the
disk will be stable against developing a bar if
\begin{equation}
  Q_{bar}=\frac{v_M}{\sqrt{M_d G / R_d}} > 1.1 \,,
\label{eq:bar}
\end{equation}
where $v_M$ is the maximum velocity of the rotation curve, and $M_d$ and
$R_d$ are the disk mass and scale length, respectively.  For the initial
live disks in our simulations at $z=1.0$, we find $Q_{bar}=0.98, 0.83,
1.03$, and 0.93 for haloes A, B, C, and D, respectively, and the value of
$Q_{bar}$ is largely independent of the initial disk orientation (see
Table~1 for the full list).  Since the same disk is used for all four
haloes, the values of $Q_{bar}$ reflect the differences in $v_M$, which in
turn depends on the masses of the haloes at $z=1.0$: $M_{vir}=1.03\times
10^{12}, 5.7\times 10^{11},1.18\times 10^{12}$, and $1.07\times 10^{12}
M_\odot$ for haloes A, B, C, and D, respectively.  Since $Q_{bar} < 1.1$,
these disks are expected to be unstable against a bar forming, which is
indeed in qualitative agreement with Fig.~\ref{fig:Images}.  Halo C has the
weakest bar since it is the most massive halo and its $Q_{bar}$ value is
closest to the threshold for stability.

In comparison, we find $Q_{bar}$ to be greater than 1.1 in our simulations
in which the disk masses are lowered by a factor of 2 and 3 (see Sec.~5
below); the corresponding $Q_{bar}$ is 1.13 and 1.23 for halo A, and 1.15
and 1.26 for halo C, respectively.  The increase in $Q_{bar}$ for
decreasing disk mass indicates stronger stability against bar formation in
lower mass disks.  These smaller disks indeed appear mostly free of bars,
as shown below in Fig.~\ref{fig:AltImage}.  The $m=2$ coefficient is also
small, with $A_2 < 0.2$ (right panels of Fig.~\ref{fig:AltShape}).
Equation~(\ref{eq:bar}) is approximate since it treats only the
self-gravity of the disk and does not consider the velocity dispersion
\citep{athanassoula08}.  For the disks studied here, however,
equation~(\ref{eq:bar}) appears to provide a reasonable approximation for
bar stability.

\begin{figure*}
\includegraphics[width=170mm]{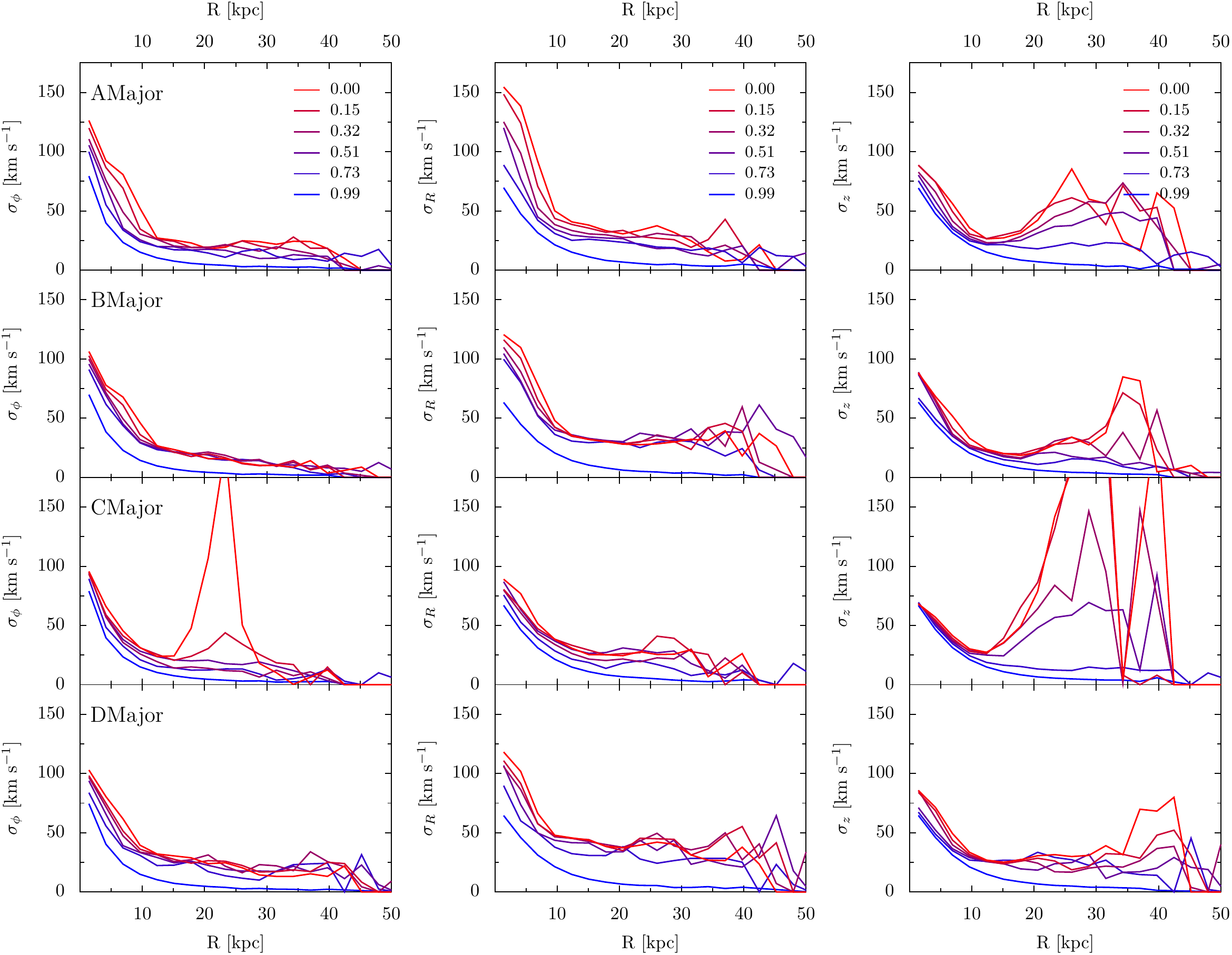}
\caption{Same as Fig.~\ref{fig:MinorVelocity} but for the four disks
  in the major orientation.}
\label{fig:MajorVelocity}
\end{figure*}

Stellar bars have been reported to dissolve in live triaxial dark matter
haloes (e.g., \citealt{berentzen06a}).  As the solid curves in
Fig.~\ref{fig:proandflat} show, our haloes at the end of the disk growth
phase at $z=1.0$ are all triaxial with fairly similar axial ratios.  For
instance, the axial ratios of the AMajor and AMinor haloes at $z=1.0$ are
1.00:0.86:0.69 and 1.00:0.87:0.62, respectively, while the CMajor and
CMinor haloes have 1.00:0.90:0.61 and 1.00:0.87:0.62, respectively.  The A
disks form bars that buckle and then continue to grow until $z=0$, but the
C disks never form appreciable bars and the $A_2$ amplitude stays nearly
constant and low.  Therefore, halo triaxiality in our simulations appears
to play a minor role in bar evolution in comparison to other factors such
as the relative disk-to-halo mass.

\subsection{Disk Heating and Velocity Profiles}

In addition to the shape of the disk, the velocity structure of the disk
evolves during the simulation.  We define the following quantity to
characterize the vertical heating of the disk.  For a cumulative radial
disk mass profile $M(R)$ and a vertical velocity dispersion profile
$\sigma_z(R)$, we use
\begin{equation}
\label{eqn:zeta}
\zeta = \frac{\int_0^{2R_d} \frac{dM}{dR} \sigma_z^2(R) dR}{\int_0^{2R_d}
   \frac{dM_0}{dR} \sigma_{z,0}^2(R) dR}
\end{equation}
where subscript ``0'' in the denominator denotes that these quantities are
evaluated for the initial redshift of the live phase ($z = 1.0$).  The
limit of integration is set to twice the scale radius, as in the
$A_2$ definition above.

The evolution of $\zeta$ with redshift is given in Fig.~\ref{fig:zeta} for
disks in the minor (top) and major (bottom) orientations.  The halo-to-halo
variation is striking.  Once again, vertical heating is negligible in the
CMajor disk, in which $\zeta$ stays at unity throughout the simulation.
The CMinor disk experiences a gradual increase of only $\sim 30$\% in
$\zeta$ between $z=1$ and 0.  By contrast, $\zeta$ jumps from 1 to 2-2.5
suddenly at $z\sim 0.5$ to 0.7 for the disks in the other three haloes.
This jump in $\zeta$ is accompanied by a dip in the bar strength $A_2$ at a
similar redshift, as a result of bar buckling (see Fig.~\ref{fig:modesum}
and previous subsection).

The complete set of profiles of the stellar velocity dispersions in three
directions about the disk plane is shown in Fig.~\ref{fig:MinorVelocity}
and \ref{fig:MajorVelocity} for the disks in the minor and major
orientation, respectively.  In each figure, the three columns show the
azimuthal (left), radial (middle), and vertical (right) components of the
stellar velocity dispersion.  Within each panel, six snapshots between
$z=1$ and 0 are plotted.

In the vertical direction in the inner $R \sim 10$ kpc of the disk, jumps in
$\sigma_z$ are clearly seen between $z\sim 0.7$ and 0.5 for the five disks
AMinor, BMinor, DMinor, BMajor and DMajor, as was shown in
Fig.~\ref{fig:zeta}.  For the other three disks, AMajor's $\sigma_z$ at
small $R$ increases more gradually without a sudden jump, while $\sigma_z$
stays nearly constant for CMajor and CMinor, which do not form a
significant bar.  This behavior is again consistent with
Fig.~\ref{fig:zeta}.

In the vertical direction at radii beyond 10 kpc, some disks show a sharp
increase in $\sigma_z$ in Figs.~\ref{fig:MinorVelocity} and
\ref{fig:MajorVelocity}.
The rise is particularly prominent for the three disks CMinor, DMinor, and
CMajor, in which $\sigma_z$ reaches $\ga 150$ km/s at $R\sim 20$ to 40 kpc.
The edge-on images in Fig.~\ref{fig:Images} indeed show that these three
disks contain an extended diffuse component of stars outside the disk
plane, spreading to tens of kpc.  This component gives rise to the sharp
increase in the mean vertical height in the three objects in
Figs.~\ref{fig:MinorStructure} and \ref{fig:MajorStructure}.  The AMajor
disk also exhibits a similar behavior to a lesser extent.  The AMinor disk,
by contrast, stays thin and cold in the outer part despite the strong bar
formed in the inner region.

In the radial and azimuthal directions, Figs.~\ref{fig:MinorVelocity} and
\ref{fig:MajorVelocity} show two phases of heating.  Between the first two
snapshots at $z=0.99$ (blue curves) and 0.73 (light purple curves), which
precedes any bar formation, $\sigma_R$ and $\sigma_{\phi}$ in the inner 10
kpc increase for all eight disks, with a larger jump for the disks in the
minor orientation.  In comparison, there is almost no vertical heating as
evidenced by the nearly identical $\sigma_z$ at these snapshots.  Similar
trends are reported in an idealized model for the halo and disk of M31, in
which the changes in $\sigma_R$ and $\sigma_\phi$ are attributed to
transient spiral features present early in the simulations
\citep{gauthier06}.

A second phase of heating in the radial and azimuthal directions can be
seen in a subset of disks in the inner $\sim 10$ kpc in the $z < 0.73$
snapshots of Figs.~\ref{fig:MinorVelocity} and \ref{fig:MajorVelocity}.  It
is tempting to attribute this phase of heating to bar activities, e.g.,
$\sigma_R$ and $\sigma_\phi$ are nearly unchanged between $z=0.73$ and 0.0
for CMinor and CMajor that have very weak bars, whereas they increase by up
to 50\% for AMinor and AMajor that have strong bars.  Fig.~6 of
\citet{gauthier06} illustrates similar heating and explains it by the
stellar bars that have formed in their runs with dark matter subhalos.  We
note, however, that the BMinor disk (and BMajor to a lesser extent)
contains a strong bar, but its $\sigma_R$ and $\sigma_\phi$ stay nearly
constant after $z=0.73$.

\begin{figure}
\includegraphics[width=84mm]{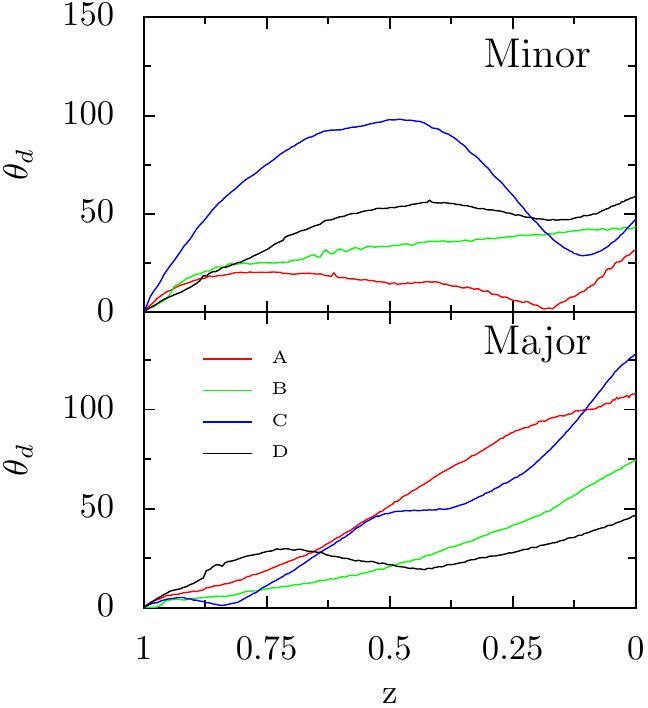}
\caption{Angle (in degrees), $\theta_d$, between the disk normal (for the
  inner 5 kpc) at a given redshift and the disk normal at $z=1.0$ for
  haloes A (red), B (green), C (blue), and D (black).  The disk normal is
  initially aligned with the minor (top) or major (bottom) axis of each
  halo at $z=1.3$.  The disks experience drastically different amounts of
  tumbling between $z=1.0$ and 0: the two C disks and AMajor disk rotate by
  more than 100 degrees, whereas the normal of the AMinor disk stays within
  $\sim 20$ degrees throughout the simulation.}
\label{fig:Orient}
\end{figure}

\subsection{Disk Reorientation and Warps}
\label{sec:Orientation}

Once the stellar disk goes live at $z=1.0$ in our simulations, the
orientation of the disk is free to change.  The outer parts of the disks
($R \ga 10$ kpc) can experience significant warps compared with the inner
parts, so we examine the two regions separately here.

For the central region, we plot in Fig.~\ref{fig:Orient} the angle,
$\theta_d$, between the disk normal computed from stars within 5 kpc at a
given redshift and the initial disk normal at $z=1.0$.  The results do not
depend sensitively on the choice of 5 kpc and are nearly identical when
stars out to $\sim 10$ kpc are included.  The CMinor, CMajor, and AMajor
disks are seen to experience the largest amounts of re-orientation by
$z=0$, but $\theta_d$ evolves differently for each case.  The CMinor disk
normal starts to tilt away from its initial direction at $z=1$ immediately,
reaching a separation of $\sim 100$ degrees at $z\sim 0.5$.  The CMajor
disk orientation, on the other hand, stays quite constant until $z\sim
0.7$, after which $\theta_d$ grows steadily and reaches $\sim 125$ degrees
at $z=0$.  AMajor resembles CMajor with $\theta_d$ reaching $\sim 110$
degrees at $z=0$.

In comparison, the central regions of the AMinor and DMajor disks
experience the least amounts of re-orientation.  The disk normal for each
case stays within $\sim 20$ degrees throughout most of the simulation, and 
rises only slightly to $\sim 30$ and 45 degrees, respectively, at $z\sim 0$.

We have also computed the relative orientation between each disk normal
shown above and its halo at a given time.  The principal axes of a halo are
determined by binning the halo particles in gravitational potential and
fitting ellipsoids to equipotential surfaces.  This procedure is repeated
at each bin in potential, which corresponds roughly to a given radius.  In
nearly all cases, the halo axes show very little twist over the inner 50
kpc, and the disk orientation (within 5 kpc) is very well aligned with one
of the halo's principal axes.  Thus, when a disk re-orients in
Fig.~\ref{fig:Orient}, it is doing so with the inner halo.
\citet{bailin05} and \citet{bailin05b} have reported similar internal
alignment in haloes' axes in N-body simulations and close alignments
between disks and the inner haloes in hydrodynamical simulations.

\begin{figure*}
\includegraphics[width=170mm]{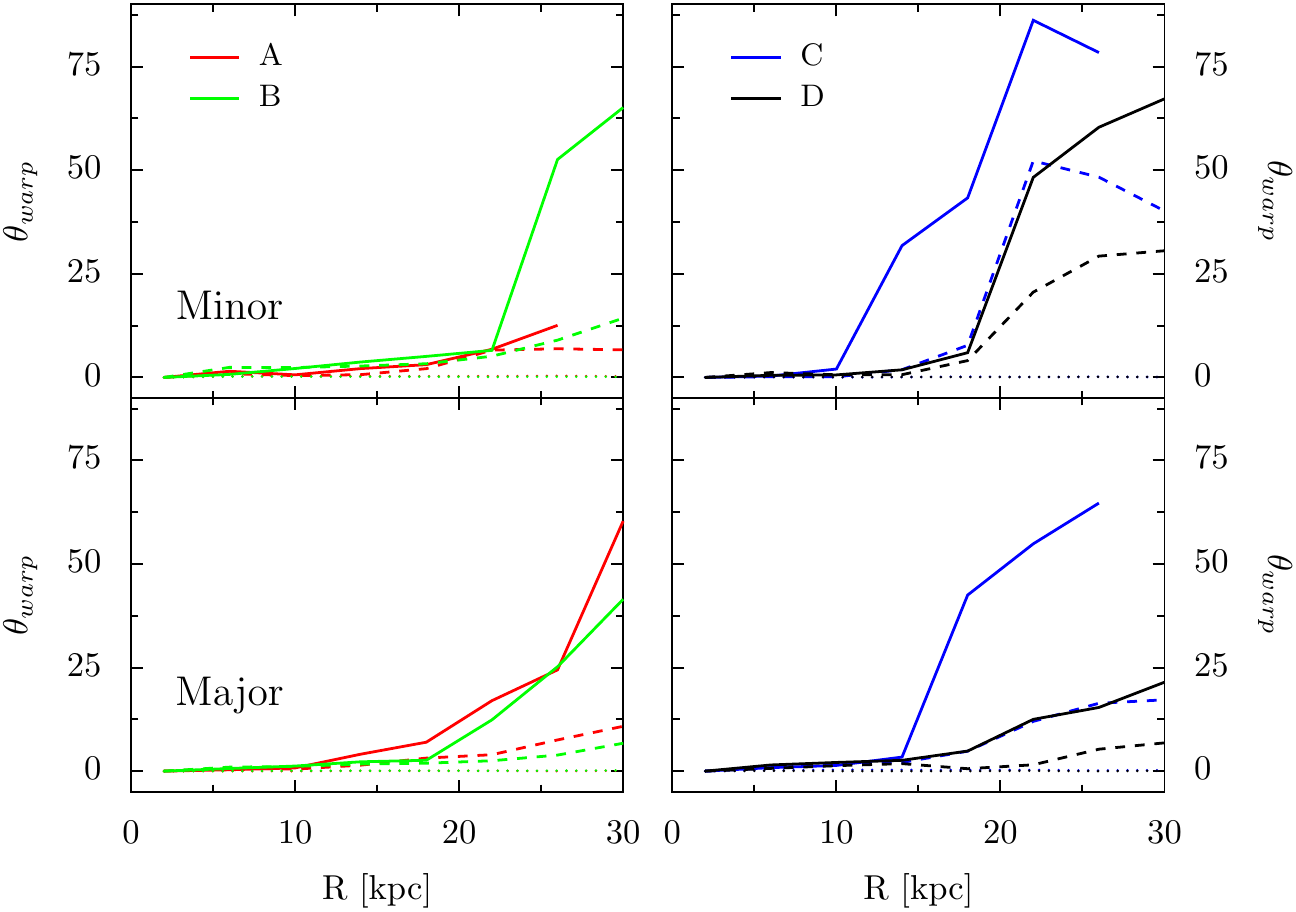}
\caption{Warp angle for the Minor (top row) and Major (bottom row) disks as
  a function of radius at three different times: $z = 1.0$ (dotted), 0.5
  (dashed), and 0.0 (solid).  The disk is flat at $z=1.0$ by construction
  so $\theta_{warp}=0$ at all radii.  The BMinor, CMinor, DMinor, and
  CMajor disks show outer warps with $\theta_{warp} > 60$ degrees at $z=0$,
  whereas the AMinor and DMajor disks stay quite coplanar out to $\sim 30$
  kpc.  }
\label{fig:TwistAngle}
\end{figure*}

Visual inspection of the stars at large radii suggests that the material is
largely in the form of rings of material concentric with the disk rather
than in a random diffuse form.  These ring-like structures are, however,
warped and offset from the inner disk.  To quantify the relative
orientation of the inner and outer parts of each disk, we plot in
Fig.~\ref{fig:TwistAngle} the warp angle, $\theta_{warp}(R)$, which is
defined to be the angle between the disk normal for the central radial bin
and the normal for the bin at radius $R$.  For each disk, this angle
measured at three redshifts, $z=1$, 0.5, and 0.0, is shown.  Initially ($z
= 1.0$; dotted lines), the disk is completely coherent by construction and
$\theta_{warp}=0$ at all radii.  As the disk evolves to $z = 0.5$ (dashed)
and $z = 0.0$ (solid), a subset of the disks is significantly beyond 10 kpc.

The trends in Fig.~\ref{fig:TwistAngle} are closely correlated with those
in Fig.~\ref{fig:Orient}: disks (and their haloes) such as CMinor, CMajor,
and AMajor that have tumbled significantly between $z=1$ and 0 develop
large outer warps by $z=0$, whereas the disks with nearly constant normals
such as AMinor and DMajor stay coplanar with $\theta_{warp}(R) \la 20$
degrees out to $R=30$ kpc.  The time of the tumbling and warping is also
correlated.  For instance, the disk that starts to tumble earliest, CMinor,
shows significant warping at the $z=0.5$ snapshot already, whereas
$\theta_{warp}$ for the CMajor disk is less than 20 degrees at $z=0.5$ but
rises to $\sim 70$ degrees at $z=0$.

The behavior above also matches the expectation from
Figs.~\ref{fig:Images}, \ref{fig:MinorStructure} and
\ref{fig:MajorStructure}: the larger the value of $\theta_{warp}$, the
more material appears out of the plane of the disk, and the larger the
increase in the disk height at large radii.

While the inner parts of the halo and disk are well aligned, the outer
region of the halo can be misaligned with the inner region
\citep{bailin05b}.  As the halo tumbles slowly over the course of the
simulation, any such misalignment can generate tidal fields that torque the
disk and cause the more fragile outer parts of the disk to warp
\citep{debattista99, dubinski09b}.  We have compared the directions of the
major and minor axes of each halo in the inner 50 kpc versus between 50 kpc
and 100 kpc at various redshifts.  We find the axes to be well aligned
without much twisting, similar to the results reported in
\citet{veraciro11}.  The warps in our simulated disks therefore appear to
be more correlated with the overall tumbling of the disk-halo system in the
cosmological simulation.
Other mechanisms have been shown to induce disturbances in the outer disk.
Perturbations from infalling satellites or cold gas, for instance, can be
sources of warping torques (e.g., \citealt{ostrikerbinney89, quinn93,
  debattista99, jiangbinney99, weinberg06, chakrabarti09, roskar10}).
The satellites generate additional tidal fields and also bring in angular
momentum that can alter the alignment of the disk and halo.  The infalling
cold gas can be strongly torqued by the hot gas halo.  These mechanisms may
act in concert on the outskirts of disks.

\section{Alternate Disk Models}
\label{sec:altdisks}

To quantify the dependence of the disk structures on the assumed disk mass,
we have performed additional simulations for halos A and C, in which the
disk mass is reduced from the fiducial value of $5\times 10^{10} M_\odot$
by a factor of 2 and 3.  The scale radius of the disk is reduced according
to the observed scaling relation $R_d \propto M_d^{1/3}$ \citep{shen03}.
These runs are labeled ``Half'' and ``Third'' in Table~1.  The change of
disk mass required the disk growing phase between $z=1.3$ and 1.0 to be
redone for each case since the dark matter haloes would be responding to
the potential of a smaller and less massive disk.

We have also performed test runs with haloes A and C in which the disk is
assumed to have an initial gas fraction of $f_g = 0.4$.  These runs are
labeled AMinorGas and CMinorGas.  The total baryonic mass is the same as in
the fiducial run, i.e. $5\times 10^{10} M_\odot$; the disk mass in stars is
therefore reduced to $3\times 10^{10} M_\odot$.  The gas has the same
radial profile as the stellar particles, but is set up to be in vertical
hydrostatic equilibrium. For simplicity it does not form stars or cool.  We
include this component to test if a gas component helps stabilize the stars
in a disk.  A more realistic simulation will turn most of the gas at $z=1$
into stars by $z=0$, thereby bracketing our results for the AMinorGas run
and the fiducial run AMinor.

\begin{figure}
\includegraphics[width=84mm]{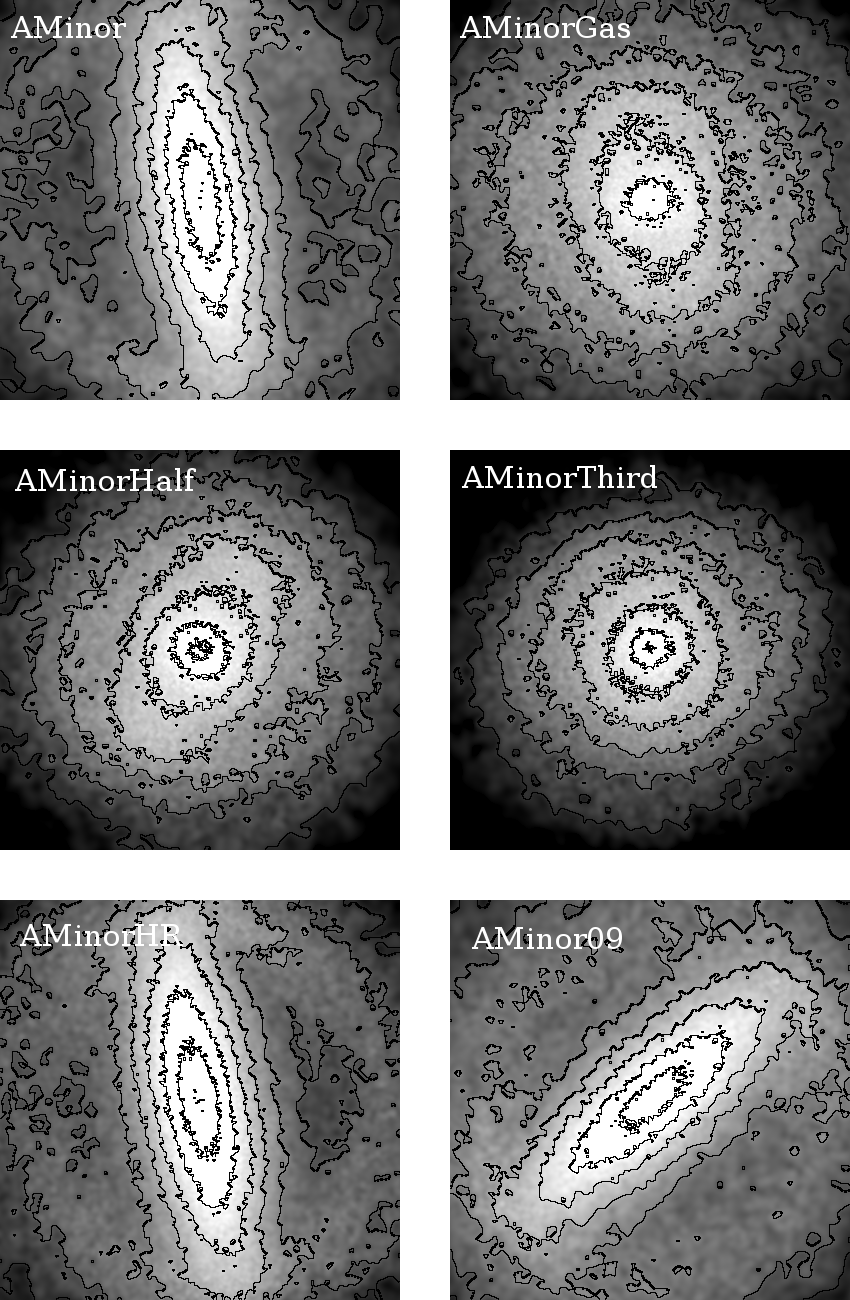}
\caption{Surface density of the stellar disk at $z=0$, viewed faced-on, for
   six disk models in halo A.  The models in the top two rows
  differ in the assumed disk masses and scale radii: $5\times 10^{10},
  3\times 10^{10}, 2.5\times 10^{10}$ and $1.67\times 10^{10} M_\odot$ for
  AMinor, AMinorGas, AMinorHalf, and AMinorThird, respectively.  Reducing
  the disk mass or placing some baryons in a gas component helps stabilize
  the disk against bar formation.   The two panels in the bottom row
    show the resulting disks in the higher resolution run `AMinorHR,' and
    in run `AMinor09' in which the disk becomes live at $z=0.9$ instead of
    1.0; the results are nearly identical to run `AMinor.'  The images are
  21 kpc on a side, and contours have been added to highlight the shape of
  the disks.  The brightness of the image is logarithmic in the surface
  density, with all four images having the same scale.  }
\label{fig:AltImage}
\end{figure}

We find large reductions in the bar strengths as the disk mass is lowered.
Fig.~\ref{fig:AltImage} shows the $z=0$ face-on surface density of the
inner 21 kpc by 21 kpc region of four different disks for halo A.  The
right column of Fig.~\ref{fig:AltShape} shows the time evolution of the
amplitude $A_2$ of the $m=2$ mode that characterizes the bar strength for
haloes A and C (see Sec 4.2).  These figures show a clear decrease in bar
strengths as the disk mass is reduced.  Halo AMinor with the fiducial mass
disk has a growing $m=2$ mode throughout the simulation, while the other
three smaller disks all have $A_2 < 0.2$.  The less massive disks in the C
haloes also have smaller values of $A_2$, but the overall decrease is
smaller as the CMinor disk shows at most a weak bar to begin with.  As
discussed in Sec 4.2, the approximate bar instability criterion given by
equation~(\ref{eq:bar}) requires $Q_{bar} < 1.1$ for bar formation.  The
value of $Q_{bar}$ for each disk at $z=1.0$ is listed in Table~1.  The
disks with the fiducial mass of $5\times 10^{10} M_\odot$ satisfy this
condition and indeed develop bars by $z=0$ (see Fig.~3).  In comparison, the
disks with reduced masses all have $Q_{bar} > 1.1$ and $A_2 < 0.2$ and show
no strong bars.

The middle column of Fig.~\ref{fig:AltShape} shows the mean vertical
distance of the stellar particles from the disk plane as a function of
radius for these same simulations at $z=0$.  Both the height and radius are
scaled to the initial disk size to aid in the comparison among disks with
different masses.  In the inner part of the disk, the bumps in $|z|$ at two
to three scale radii due to the bars in the fiducial disks are absent in
the lower mass disks, consistent with the lack of bars in the latter.
In the outer part of disk, the less massive disks are more easily brought
out of the disk plane, leading to streamers that are originally
closer to the center of the disk.

In contrast to the strong dependence of bar strengths on disk mass, we find
the less massive disks to have similar effects as the fiducial disk on the
dark matter halo between $z=1.3$ and 1.0 as the rigid disk is introduced.
The left column of Fig.~\ref{fig:AltShape} shows that the axial ratio $b/a$
for both haloes A and C is driven towards unity, while $c/a$ is largely
preserved as the rigid disk is brought to its final mass, nearly
symmetrizing the halo in the disk plane.  The effects on the halo shape do
not weaken as the final disk mass is reduced by a factor of three.  The
axial ratios are shown for a particular equipotential surface, chosen to be
about 5 kpc from the center of the disk.

The less massive disks also reorient themselves during the simulations.
All four AMinor disks have similar reorientation histories, showing at most
a 12 degree offset between the disk orientations. The alternate CMinor
runs, however, have more divergent reorientation histories.  The angle
between the original CMinor disk normal and the disk normal in the
alternate CMinor runs is typically around 30 degrees, rising to as high as
60-80 degrees temporarily near $z \sim 0.7$.

Our simulations that include gas show drastically reduced signatures of
bars, and the associated heating.   This reduction matches the observed
  decrease in the likelihood of hosting a bar in galaxies with increasing
  gas fraction \citep{masters12}.  The strong effect of this massive gas
component is expected because the gas offers support against self-gravity
via its pressure.  The extra support provided in these simulations is
likely an overestimate, as realistic gas cools and forms stars, but the
ability of a simple gas model to suppress the bar points to the need to
include a realistic gas component.  The lack of a strong bar also means
that the subsequent bar buckling cannot occur and the central thickness of
the disk is smaller at the end of the simulation.

All of our simulations start with a rigid disk potential at $z=1.3$
  and transition to a live disk at $z=1.0$.  To test if any of our results
  is sensitive to the choice of $z=1.0$, we have run a simulation of the
  AMinor halo in which the rigid disk becomes live at a later time of $z =
  0.9$ (`AMinor09' in Table~\ref{table:simparm}).  The qualitative features
  of this run are in good agreement with the AMinor run (Fig.~12): a bar
  forms and eventually develops an X-shape, the disk heats vertically, and
  the disk reorients and shows signs of material out of the plane of the
  disk.

To assess the degree of convergence of these results with the resolution of
the stellar disk, we have run a disk in the AMinor orientation with
$5\times 10^5$ particles instead of $2\times 10^5$, labeled AMinorHR in
Table~\ref{table:simparm}.  The results of this simulation, e.g., the
values of $A_2$ and $\zeta$, agree to within 5\% with the run at our
fiducial resolution (see also Fig.~12).  A similar degree of convergence is
reported in the detailed study of \citet{dubinski09}.  Quantities such as
the bar strength, pattern speed, and halo response to disk in their
simulation with $10^6$ dark matter particles and $1.8\times 10^5$ disk
particles are very similar to those in runs with 10 and 100 times more
particles.

\section{Summary and Discussion}
\label{sec:conclusion}

We have presented the results of simulations of live stellar disks in the
fully cosmological setting of the Aquarius simulations of dark matter haloes 
similar in mass to that of the
Milky Way.  Our simulations take place in two phases.  We first
allow the dark matter halo to respond adiabatically to a growing rigid disk
potential between $z=1.3$ and 1.0.  A live disk of stellar particles is
then inserted and evolved self-consistently with the dark matter particles
from $z=1.0$ to today.  We have inserted disks in four Aquarius haloes (A,
B, C, and D), with the disk normal initially aligned with the minor or major 
axis of the halo at $z=1.3$.  The main results in this paper are drawn
from these eight simulations; several additional runs are also performed to
test numerical convergences and to assess the impact of different disk models
(Table~1).

The disks have a strong effect on the shapes of the inner haloes.  The
haloes are approximately prolate in the inner $\sim 50$ kpc at $z=1.3$.  
As a rigid disk potential is added, the inner haloes evolve
to become oblate by $z=1.0$ (Figs.~\ref{fig:HaloImages} and
\ref{fig:proandflat}).  This change occurs regardless of the disk
orientation but it occurs for different reasons for the two orientations.
When the disk normal is initially aligned with the halo's minor axis, the
two longer axes that define the disk plane become more symmetrized by $z=1.0$
due to the axisymmetric disk potential. The inner halo thus becomes nearly
oblate.  When the disk normal is initially aligned with the halo's major
axis, however, the original major axis shrinks and becomes the minor axis
by $z=1.0$.  This also results in an oblate inner halo, where the disk normal
aligns with the minor axis.  Thus when the live disk simulation
starts at $z=1.0$, the disk normal is aligned with the halo's minor axis in
both of our ``Major'' and ``Minor'' runs.

\begin{figure*}
\includegraphics[width=170mm]{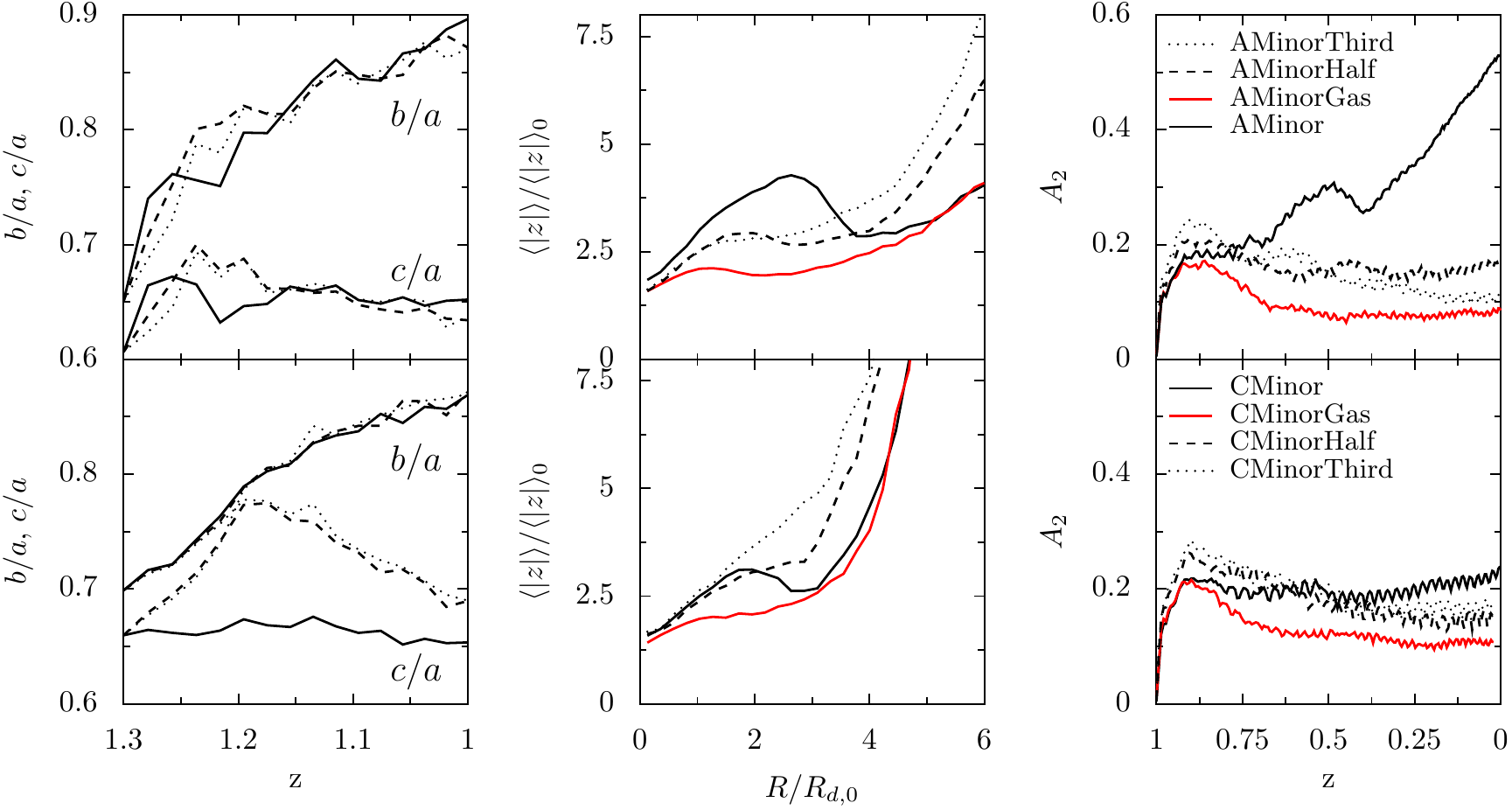}
\caption{Comparison of four disk models for halo A (top row) and halo C
  (bottom row).  The four models differ in the assumed disk mass: $5\times
  10^{10}, 3\times 10^{10}, 2.5\times 10^{10}$ and $1.67\times 10^{10}
  M_\odot$ for runs labeled Minor, MinorGas, MinorHalf, and MinorThird,
  respectively.  \emph{Left:} The axial ratios of the dark matter haloes
  between $z=1.3$ and 1.0 hardly depend on the disk mass.  \emph{Middle:}
  Mean vertical distance $|z|$ of the stellar particles from the disk plane
  as a function of disk radius at $z=0$.  Among these runs, only the AMinor
  and CMinor disks (black solid curves) form stellar bars, which produce a
  bump in $|z|$ at a radius of $\sim 2$ - 3 $R_d$.  \emph{Right:} Bar
  strengths measured by the $m=2$ mode amplitude $A_2$ as a function of
  redshift.  Reducing the disk mass stabilizes the disk against bar
  formation.  Note that the runs including gas are not included in the left
  panels because before the disk goes live, their evolution is identical to
  the original `Minor' runs.}
\label{fig:AltShape}
\end{figure*}

For the live disks, we find striking variations in the evolution of their
spatial and velocity structures between $z=1.0$ and 0.  The A, B, and D
disks form strong bars that thicken and heat the stellar particles in the
inner $\sim 10$ kpc, whereas the C disks are nearly bar-free and remain
thin and vertically cold at $R \la 10$ kpc.  Five of the bars experience
buckling instabilities between $z\sim 0.7$ and 0.5, resulting in a sudden
increase in the vertical velocity dispersion $\sigma_z$ and an accompanying
dip in the amplitude $A_2$ of the $m=2$ Fourier mode of the surface density
(Figs.~\ref{fig:modesum} and \ref{fig:zeta}).  These bars are not destroyed
by the buckling but instead continue to grow from $z\sim 0.5$ until the
present day.

The behavior in the outer parts of the disks also varies greatly from disk to
disk.  Among the eight disks, the CMajor and CMinor disks show the largest
amount of ring-like material out of the plane of the disk.
Similar components are also clearly seen in the AMajor and DMinor disks
(Fig.~\ref{fig:Images}).  These stars lie in warped planes at the
outskirts, where the disk normal can twist by up to $\sim 80$ degrees from
the disk normal in the inner parts (Fig.~\ref{fig:TwistAngle}).  The stars in
this subset of disks greatly increase the mean vertical height
(Figs.~\ref{fig:MinorStructure} and \ref{fig:MajorStructure}) and the vertical
velocity dispersion (Figs.~\ref{fig:MinorVelocity} and \ref{fig:MajorVelocity}) at
$R\ga 10$ kpc, relative to the plane of the inner disk.

We have found a strong correlation between the severity of the outer warps
and the amount of tumbling that the disk-halo system has undergone
during the course of the simulation, i.e., the more tumbling, the
larger the warps (Figs.\ref{fig:Orient} and \ref{fig:TwistAngle}).  In
addition, the halo and the disk are very well aligned and appear to tumble
together.  These trends suggest that the haloes undergoing larger tumbling
experience larger tidal fields that torque the fragile outer parts of
the disks.

Reducing the disk mass helps stabilize the disk against bar formation.  The
strong bar in the fiducial run for halo A is mostly absent when the disk
mass is lowered from $5\times 10^{10} M_\odot$ to $2.5\times 10^{10}
M_\odot$ and $1.67\times 10^{10} M_\odot$ (Figs.~\ref{fig:AltImage} and
\ref{fig:AltShape}).
Placing 40\% of the original disk mass into a coaligned gas 
disk also reduced the bar drastically.  The gas was not allowed to cool or 
form stars, thus artificially enhancing its stabilising effect. The purely 
stellar disk and the disks with an initial gaseous component can be thought of 
as bracketing the results for a more realistic gas model.

The disks presented in our simulations share many features with observed
disk galaxies, although a detailed comparison would require a larger
statistical sample as well as the inclusion of gas in the simulations.  We
can nonetheless make some broad comparisons within the limited statistics.
Six of the eight disks in our simulations easily form bars at $z\sim 0.7$
to 0.5 and the bars remain stable until the present-day.  Similarly, as
many as 2/3 of the nearby disk galaxies have been observed to host a bar
(\citealt{marinova07} and references therein).  Though the evolution of the
bar fraction with redshift is still a subject of ongoing debate (e.g.,
\citealt{jogee04,sheth08}), the fraction of barred disks is non-negligible
out to $z \sim 1$.   Our simulations that include gas show
  significantly reduced signatures of bars.  This reduction matches the
  observed decrease in the likelihood of hosting a bar in galaxies with
  increasing gas fraction \citep{masters12}.

In addition to the prevalence of bars, as discussed in Sec~4.1, the
X-shaped structures or peanut-shaped bulges that have been seen in many
nearby edge-on spiral galaxies are also produced in our simulated disks
(Fig.~\ref{fig:Images}).  One conspicuous feature in a subset of our disks
is the misalignment between the orientation of the inner and outer regions
of the disk (Fig.~\ref{fig:TwistAngle}).  By eye, the structures resemble
observed warps in disk galaxies. These are quite common in the local
universe, as seen in the stellar light and (particularly) in neutral
hydrogen in both our own Galaxy and in other nearby systems (e.g.,
\citealt{kerr57, sancisi76, reshetnikov98, garciaruiz02, levine06,
  kruitfreeman11} and references therein).

The circular velocity profiles, $V_{\rm circ}(r)=\sqrt{GM(r)/r}$, of
  disks A, C, and D have broad peaks reaching 220 to 260 km/s at radii
  between 5 and 10 kpc and decline gently outward.  This is similar to the
  observed profile for the Milky Way \citep{xue08}.  The virial mass of
  halo B is lower than that of A, C, and D by nearly a factor of two and
  therefore has a lower $V_{\rm circ}(r)$.  For the velocity dispersions,
  the general trend of $\sigma_R > \sigma_\phi > \sigma_z$ in the inner
  $\sim 10$ kpc is consistent with the Milky Way.  At 8 kpc, the coldest
  disk (CMajor) has $(\sigma_R, \sigma_\phi, \sigma_z) \approx (45, 40,
  35)$ km/s at $z=0$, somewhat higher than (38, 26, 19) km/s found for red
  stars in the solar neighbourhood \citep{aumer09}.  While some disks
  experience noticeable heating, the vertical structure of the CMajor disk
  stays quite constant within $\sim 10$ kpc between $z=1.0$ and 0.0
  (Fig.~5), similar to a sample of edge-on disk galaxies surveyed by
  Spitzer \citep{comeron11}.

The focus of this paper is on the gravitational interactions between live
stellar disks and their host dark matter haloes in realistic cosmological
simulations.  We have intentionally left out discussions of the dark matter
subhaloes, which are well resolved in zoom-in simulations such as the
Aquarius haloes.  As discussed in Sec.~4, infalling satellite galaxies,
dark matter substructures, and cold gas clumps have been shown in prior
studies to increase disk thickness, induce bar formation, and cause
disturbances in the outer disks such as warps.  On the other hand, strong
bars and severe warps arise in different disks in our simulations and do
not appear to be directly related.  Moreover, we have demonstrated that the
bar strengths depend sensitively on the disk-to-halo mass ratio, and that
the outer warps are strongly correlated with the overall rotation of the
disk-halo system.  Therefore, it is presently unclear if dark matter
subhaloes play a major role in producing any of the disk features reported
in this paper.  The disks can in turn alter the fate of massive subhaloes
that pass through them \citep[e.g.,][]{donghia10}.  The magnitudes of all
these effects depend sensitively on the orbits, masses, and concentrations
of the subhaloes, in particular, on whether a massive subhalo can survive
in the vicinity of the disk.  We plan to quantify the impact of subhaloes
on the spatial and kinematical structures of the disks and the influence of
the disks on the survivability of the subhaloes in a future work.

\section*{Acknowledgments}

We thank Michael Aumer, Chris McKee and Michael Boylan-Kolchin for useful 
discussions and suggestions, and Onsi Fakhouri for contributions to the 
earliest part of this project.  Support for C.-P.M is provided in part by 
NASA through grant NNX11AI97G and grant HST-AR-12140.01-A from the Space 
Telescope Science Institute, and by the Miller Institute for Basic Research 
in Science, University of California, Berkeley.  This research used resources 
of the National Energy Research Scientific Computing Center, which is supported 
by the Office of Science of the U.S.  Department of Energy under Contract
No. DE-AC02-05CH11231, and the Henyey cluster at UC Berkeley supported by
NSF AST grant 0905801.

\bibliographystyle{mn2e} 
\bibliography{thesis}
\label{lastpage}

\end{document}